\documentclass[twocolumn, showpacs, superscriptaddress, nofootinbib]{revtex4-1}
\usepackage[utf8]{inputenc}
\usepackage[english]{babel}
\usepackage{graphicx}
\usepackage{float}
\usepackage{amsmath}
\usepackage{mathtools}			

\begin{document}

\title{Statistical description of mobile oscillators in embryonic pattern formation}

\author{Koichiro Uriu}
\affiliation{School of Life Science and Technology, Tokyo Institute of Technology, 2-12-1, Ookayama, Meguro-ku
Tokyo 152-8550, Japan}
\affiliation{Graduate School of Natural Science and Technology, Kanazawa University, Kakuma-machi, Kanazawa 920-1192, Japan}
\author{Luis G. Morelli}		
\affiliation{Instituto de Investigaci\'{o}n en Biomedicina de Buenos Aires (IBioBA) -- CONICET/Partner Institute of the Max Planck Society, Polo Cient\'{i}fico Tecnol\'{o}gico, Godoy Cruz 2390, Buenos Aires C1425FQD, Argentina}

\date{This manuscript was compiled on \today}

\begin{abstract}
Synchronization of mobile oscillators occurs in numerous contexts, including physical, chemical, biological and engineered systems.
In vertebrate embryonic development, a segmental body structure is generated by a population of mobile oscillators.
Cells in this population produce autonomous gene expression rhythms, and interact with their neighbors through local signaling. 
These cells form an extended tissue where frequency and cell mobility gradients coexist.
Gene expression kinematic waves travel through this tissue and pattern the segment boundaries.
%
It has been shown that oscillator mobility promotes global synchronization. 
%
%
%
However, in vertebrate segment formation, mobility may also introduce local fluctuations in kinematic waves and impair segment boundaries.
%
%
Here we derive a general framework for mobile oscillators that relates local mobility fluctuations to synchronization dynamics and pattern robustness.
%
We formulate a statistical description of mobile phase oscillators in terms of a probability density.
%
%
We obtain and solve diffusion equations for the average phase and variance, revealing the relationship between local fluctuations and global synchronization in a homogeneous population of oscillators.
Analysis of the probability density for large mobility identifies a mean-field transition, where locally coupled oscillators start behaving as if each oscillator was coupled with all the others. 
We extend the statistical description to inhomogeneous systems to address the gradients present in the vertebrate segmenting tissue.
The theory relates pattern stability to mobility, coupling and pattern wavelength. 
%
%
%
%
%
The general approach of the statistical description may be applied to mobile oscillators in other contexts, as well as to other patterning systems where mobility is present.
\end{abstract}
\maketitle



\section{Introduction}
Synchronization of interacting oscillators is ubiquitous in nature~\cite{sync, pikovsky}. 
Interactions between oscillators are often between local neighbors.
With local interactions, oscillators movement affects synchronization dynamics, since movement causes the exchange of interacting neighbors.
Mobile oscillators can be found in diverse contexts, including chemical~\cite{taylor09}, biological~\cite{uriu17b} and technological systems~\cite{buscarino06}.

\begin{figure}[t]
\includegraphics[width=8.4cm]{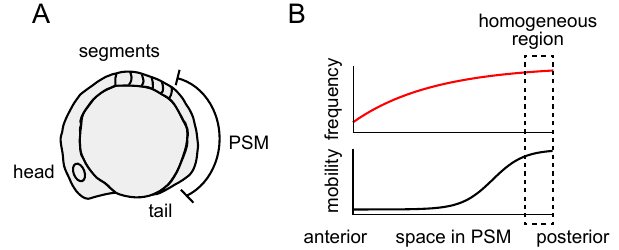}
\centering
\caption{Vertebrate segmenting tissue.
(A) Zebrafish embryo as an example of vertebrate segment formation.
Cell mobility in zebrafish presomitic mesoderm (PSM) has been quantified in detail~\cite{lawton13,uriu17a,mongera18}.
(B) Spatial gradients of oscillation frequency and cell mobility rate along the anterior-posterior axis of the PSM. In the posterior PSM, both frequency and cell mobility rate are spatially uniform.
}
\label{fig:segtissue}
\end{figure}

The problem of mobile coupled oscillators arises naturally in the context of oscillatory gene expression patterns during vertebrate body segmentation, where local intercellular interactions occur together with neighbor exchange due to cell movements~\cite{uriu14b}.
%
%
%
%
Vertebrate body segments form rhythmically from anterior to posterior, while embryos elongate (Fig.~\ref{fig:segtissue}A).
These segments bud off from an unsegmented tissue called presomitic mesoderm (PSM).
Cells in this unsegmented tissue possess a biochemical oscillator where negative feedback loops in gene regulations drive the oscillation of mRNA and protein concentrations~\cite{lewis03, schroter12}.
Neighboring cells interact using ligand and receptor proteins expressed on cell membrane.
By this intercellular interaction, phase information is transferred, leading to synchronization between neighboring cells~\cite{jiang00,horikawa06,riedel07,delaune12}.
In addition, the frequency of the cellular oscillators changes along the embryonic axis (Fig.~\ref{fig:segtissue}B)~\cite{shih15,rohde24}.
The difference in oscillation frequency results in traveling phase waves across the anterior-posterior axis of the segmenting tissue, as revealed by protein~\cite{soroldoni14} and mRNA expression~\cite{eck24}.
A new segment boundary is determined when a wave comes to the anterior part of the tissue ~\cite{oginuma10, yabe23}.
Furthermore, there is a cell mobility gradient in the segmenting tissue, with cells in the posterior part exchanging neighbors more often than those in the anterior (Fig.~\ref{fig:segtissue}B)~\cite{lawton13, uriu17b, mongera18}.
Thus, segment formation relies on spatial gene expression patterns that emerge from a frequency gradient in the presence of a cell mixing gradient.
%
To form normal segment boundaries, robustness of traveling waves is crucial~\cite{uriu21}.
Thus, oscillatory gene expression patterns in vertebrate body segment formation are a model system to address the effects of cell movements on pattern formation~\cite{pourquie11, oates12, hubaud14}.

Various modeling approaches have been developed to analyze mobile oscillators~\cite{ghosh2022}.
Previous studies investigated the effects of random motion~\cite{frasca08,uriu10a,fujiwara11,uriu13,majhi17,levis17,majhi19,paulo21,li22}, anomalous diffusion~\cite{grossmann16}, active fluid dynamics~\cite{banerjee17} and collective movement~\cite{uriu14a}.
Movement has also been described as changes in coupling network topology between oscillators~\cite{skufca04,zhou16,anwar22,majhi22}.
In addition, recent studies analyzed pattern formation driven by movement that depends on oscillators phase~\cite{tanaka07,okeeffe17,levis19,sar22,okeeffe22,ceron23}.
These previous studies indicated that mobility facilitates global synchronization.
However, a complete picture for how mobility contributes to synchronization is still lacking.
Neighbor exchange induces local phase fluctuations in spatial patterns~\cite{uriu13}, 
but how these fluctuations relate to synchronization remains unknown.
%
Furthermore, previous work has not considered the effects of mobility in inhomogeneous systems, where patterns and mobility gradients can occur together.
%
%
This is especially important for biological patterns, where cells in tissues often mix while keeping the integrity of spatial gene expression patterns~\cite{fulton22}.

In this study, we derive a general statistical description of mobile phase oscillators motivated by the vertebrate segmentation clock.
We first consider spatially homogeneous space, and reveal relations between local phase fluctuation and global synchronization.
Then we extend the theory to spatially inhomogeneous space, and show how the interplay of coupling, mobility and pattern wavelength determines pattern robustness.
%
%
%
%
%

\section{Mobile phase oscillators in homogeneous space}
%
%
In the posterior region of the segmenting tissue, frequency and mobility gradients are mostly uniform (Fig.~\ref{fig:segtissue}A, B).
Therefore, we first address the problem of mobile oscillators in homogeneous space.
We consider a population of $N$ mobile phase oscillators in a one-dimensional lattice of length $L$~\cite{uriu13, petrungaro17, petrungaro19}. 
%
%
%
Lattice sites have spatial positions $x_i=i \Delta x$ and the phase of the oscillator at site $i$ is $\theta_i(t)$, $i=0,\ldots,N$.
%
%
%
Oscillators have an identical frequency $\omega$ and interact with each other,
\begin{align}
\dot\theta_i(t) = \omega + \kappa \sum_{j=0}^{N} {C_{ij} \sin\left(\theta_j(t) - \theta_i(t)\right)},
\label{eq.phase_cont}
\end{align}
where $\kappa$ is the coupling strength and $C_{ij}$ is a coupling kernel depending on the distance between sites $i$ and $j$, that accounts for the spatial range of the interaction. 
Coupling depends on the phase difference between two lattice sites and we use a sinusoidal coupling function~\cite{kuramoto}.
\begin{figure}[t!]
\includegraphics[width=8.3cm]{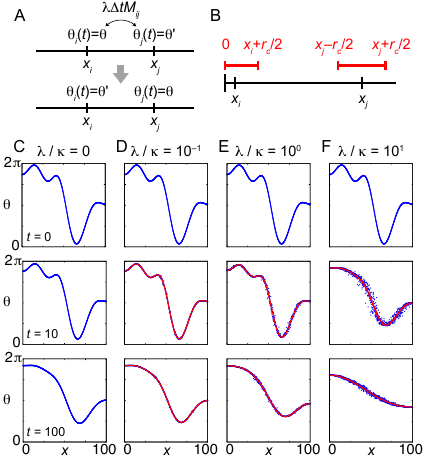}
\centering
\caption{Spatial phase patterns of mobile oscillators.
(A) Exchange of phase values between two lattice sites.
(B) Uniform coupling kernel and open boundary condition.
(C)-(F) Time evolution of a spatial phase pattern, starting from a superposition of spatial Fourier modes (Appendix A).
Snapshots at $t = 0$, $10$, and $100$ are shown.
Blue dots indicate phases from an individual realization.
Red lines indicate the average phase over $10^3$ realizations of simulations from the same initial condition.
(C) $\lambda/\kappa = 0$, (D) $10^{-1}$, (E) $10^{0}$ and (F) $10^{1}$ with $\kappa = 1$ and $\omega = 0$.
We use uniform coupling and mobility kernels with range $r_c = r_m = \sqrt{12}$ for later convenience (Appendix A).
The number of oscillators is $N=1001$, separated by $\Delta x = 0.1$ in a domain of length $L = 100$.
}
\label{fig:x_phase}
\end{figure}

To describe the mobility of oscillators, we consider the exchange of phase values between two lattice sites $i$ and $j$ at random times (Fig.~\ref{fig:x_phase}A).
This random exchange is an homogeneous Poisson process with rate $\lambda$~\cite{uriu13, petrungaro17, petrungaro19}.
%
%
The pair of oscillators that exchange positions is determined by a mobility kernel $M_{ij}$.
Namely, the probability of having a phase exchange between two lattice sites at $i$ and $j$ in a short time interval $\Delta t$ is $\lambda \Delta t M_{ij}$ (Fig.~\ref{fig:x_phase}A).
%
%
We assume the mobility kernel is symmetric $M_{ij} = M_{ji}$ and normalized $\sum_{j=0}^{N}{M_{ij}} = 1$ in the bulk.
This spatially symmetric mobility kernel represents random motions of oscillators.

We use open boundary conditions both for coupling and mobility (Fig.~\ref{fig:x_phase}B; Appendix A), 
so twisted states~\cite{wiley06, peruani10} do not occur, and global synchronization is approached independently of initial conditions~\cite{uriu13, petrungaro19}.
%
%
We illustrate the effect of mobility on transient synchronization dynamics with numerical simulations of Eq.~(\ref{eq.phase_cont}) with spatially uniform coupling and mobility kernels (Appendix A). 
For higher mobility rates $\lambda/\kappa > 1$, phase differences decrease faster (Fig.~\ref{fig:x_phase}C-F), consistent with previous reports~\cite{uriu13}.
We find that the ensemble average of phase values is a good approximation of individual simulations, and is smooth even for high mobility (red lines in Fig.~\ref{fig:x_phase}D-F).
These observations motivate a statistical description of mobile phase oscillators in terms of a probability density.

\section{Statistical description in homogeneous space}
\subsection{Probability density}
We introduce a probability density $\rho(\theta, x, t)$ for observing phase $\theta$ at position $x$ at time $t$ in an ensemble of simulations.
In a small time interval, there are two possible events: either 
(1) no exchange occurs and phases evolve according to coupled oscillator dynamics, or 
(2) a phase exchange takes place between positions $x$ and $x'$.
We derive conditional probabilities for these two possibilities, and use these to obtain the time evolution of $\rho(\theta, x, t)$
\begin{align} \label{eq.master_preexpan}
\frac{\partial \rho(\theta, x,t)}{\partial t} = &- \frac{\partial}{\partial \theta} \left[ F(\theta,x, t) \rho(\theta, x, t) \right]  \\
& + \lambda \left[ \int_0^L dx' M(x'-x) \rho(\theta, x', t) - \rho(\theta, x, t) \right], \nonumber
\end{align}
with
\begin{align}
F\left(\theta, x,t\right) = \omega \nonumber + \kappa \int_0^{2\pi} d\theta' \int_0^L dx' C(x'-x) \sin \left(\theta' - \theta \right) \\ 
\times \rho(\theta', x', t | \theta, x, t),
\label{eq.F}
\end{align}
where $\rho(\theta', x', t | \theta, x, t)$ is the conditional probability of observing phase value $\theta'$ at position $x'$ at time $t$ given the phase value $\theta$ at $x$, and we introduced mobility and coupling normalized kernels $M(x)$ and $C(x)$ in continuum space (Appendix A, B).
%
%
The first term in Eq.~(\ref{eq.master_preexpan}) describes the Liouville deterministic change in probability density due to coupled oscillator dynamics.
%
%
If we considered additional sources of noise in Eq.~(\ref{eq.phase_cont}), such as dynamic phase fluctuations, this first term would result in a Fokker-Planck equation with an additional diffusion term.
The second term in Eq.~(\ref{eq.master_preexpan}) is the master equation part describing the stochastic phase exchanges.
Together, they constitute a Chapman-Kolmogorov equation for $\rho(\theta, x, t)$~\cite{gardiner09}.

\subsection{Average phase}
To examine how mobility influences phase dynamics, we set $\omega=0$ and introduce a local average phase ${\bar \theta(x, t)} \equiv \int_{0}^{2\pi} d\theta \, \theta \rho(\theta, x, t)$.
%
%
Since the phase itself is not a periodic function, this direct average is meaningful if the phase variance is small and phase values are constrained within a $2\pi$ interval. 
%
%
%
We ensure this setting $\rho(0,x,t) = \rho(2\pi,x,t) = 0$ at any $t$, so that the support of the probability density is bounded in this interval.
For example, any initial condition that is a superposition of Fourier modes in the interval $(0,2\pi)$ satisfies these conditions.
The advantage of this definition is that we can obtain a closed equation for average phase dynamics from Eq.~(\ref{eq.master_preexpan}).
We take the time derivative
\begin{equation}
\frac{\partial {\bar \theta}(x, t)}{\partial t}  = \int_{0}^{2\pi} d\theta \, \theta \, \frac{\partial \rho(\theta, x, t)}{\partial t}  \, ,
\label{eq.der_ave_phase}
\end{equation}
and substitute Eq.~(\ref{eq.master_preexpan}) into Eq.~(\ref{eq.der_ave_phase}), truncating higher order terms $O(|x' - x|^4)$ and  $O(|\theta' - \theta|^3)$.
The resulting diffusion equation for the average phase in bulk is (Appendix B-D):
\begin{equation}
\frac{\partial {\bar \theta}(x, t) }{\partial t}  =  \frac{{\bar \kappa} + {\bar \lambda} }{2} \,  \frac{\partial^2 {\bar \theta}(x, t)}{\partial x^2},
\label{eq.pde_ave}
\end{equation}
where ${\bar \kappa} = \kappa C_2$ and ${\bar \lambda} = \lambda M_2$, with $C_2$ and $M_2$ the second moments of the coupling and mobility kernels, respectively (Appendix A).
%
%
Eq.~(\ref{eq.pde_ave}) indicates that coupling and mobility have the same effect on the average phase, contributing to an effective diffusion coefficient 
$({\bar \kappa}+{\bar \lambda})/2$.
%
%
%
Since phase diffusion makes the phase homogeneous across space, an increase in $\lambda$ accelerates the reduction of spatial phase differences as shown in Fig.~\ref{fig:x_phase}C-F.
\begin{figure}[t]
\centering
\includegraphics[width=8.3cm]{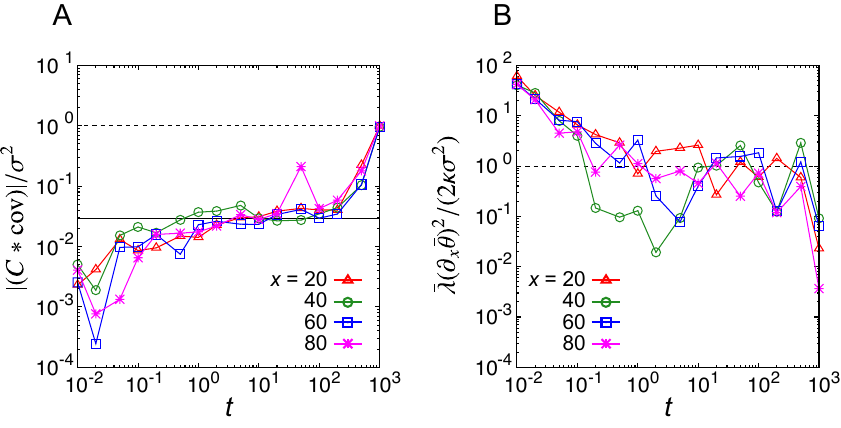}
\caption{
Comparison of covariance convolution and phase gradient square with variance.
(A), (B) Time evolution of (A) ratio of covariance convolution $(C*\mbox{cov})$ to variance and (B) ratio of phase gradient square to variance.
Solid line in (A) indicates $\Delta x/r_c$.
$\lambda / \kappa = 10^{1}$ with the same initial condition as in Fig.~\ref{fig:x_phase}.
The values at $x = 20$ (red triangles), $40$ (green circles), $60$ (blue squares) and $80$ (magenta crosses) are plotted in each panel.
Lines between points are visual guides.
Uniform coupling and mobility kernels are used in simulations as in Fig.~\ref{fig:x_phase}.
In the lattice model Eq.~(\ref{eq.phase_cont}) with a uniform coupling kernel (Appendix A), we obtain $(C*\mbox{cov}) = \Delta x \sigma^2(x,t)/r_c$.
%
%
For $t \gg 1$, phase differences between oscillators are already small and the population is about to reach complete synchronization.
This causes an increase in correlations between $x$ and $x'$, and the ratio $(C*\mbox{cov})/\sigma^2$ gradually approaches $1$ (dotted line in A).
}
\label{fig:S1}
\end{figure}

\subsection{Phase variance}
Next, we introduce the phase variance $\sigma^2 (x,t) \equiv \int_{0}^{2\pi}d\theta \left( \theta - \bar{\theta}(x, t) \right)^2 \rho(\theta, x, t)$.
We take the time derivative of $\sigma^2$ and substitute Eqs.~(\ref{eq.master_preexpan}) and (\ref{eq.pde_ave}),
truncating higher order terms to obtain the evolution of variance at the lowest order (Appendix E)
\begin{align}
\frac{\partial \sigma^2 (x,t)}{\partial t} = 2\kappa \int_0^L dx' &C(x'-x) \mathrm{cov}(x,x',t) -2\kappa \sigma^2(x,t) \nonumber \\&+ \frac{{\bar \lambda}}{2} \frac{\partial^2 \sigma^2 (x,t)}{\partial x^2} + {\bar \lambda} \left( \frac{\partial {\bar \theta}(x,t)}{\partial x} \right)^2,
\label{eq.var_full}
\end{align}
with the phase covariance
%
\begin{align}
\mathrm{cov}(x,x',t) \equiv  \int_0^{2\pi} d\theta \int_0^{2\pi}  d\theta' (\theta - {\bar \theta}(x, t)) (\theta' - {\bar \theta}(x', t)) \nonumber \\ \times \rho(\theta',x',t; \theta,x,t).
\label{eq.cov}
\end{align}
Eq.~(\ref{eq.var_full}) is not a closed equation for the variance since it involves the covariance convolution $(C*\mathrm{cov})(x,t) \equiv \int^L_0 dx' C(x'-x) \mathrm{cov}(x,x',t)$.
At $x' = x$, $\mathrm{cov}(x,x,t) = \sigma^2(x,t)$.
Due to random mobility, we may assume $\rho(\theta',x',t; \theta,x,t) \approx \rho(\theta',x',t) \rho(\theta,x,t)$ for $x' \neq x$.
This results in $\mathrm{cov}(x,x',t) \approx 0$, and consequently $(C*\mathrm{cov})(x,t) \approx 0$, consistent with numerical simulations (Fig.~\ref{fig:S1}A). 
Neglecting the covariance term, we obtain the closed equation for the variance,
\begin{align}
\frac{\partial \sigma^2 (x,t)}{\partial t} \approx - 2 \kappa \sigma^2(x,t) + \frac{{\bar \lambda}}{2} \frac{ \partial^2 \sigma^2(x,t)}{\partial x^2}  + {\bar \lambda} \left(\frac{\partial {\bar \theta}(x,t)}{\partial x}\right)^2,
\label{eq.varapprox}
\end{align}
together with the average phase Eq.~(\ref{eq.pde_ave}).

Eq.~(\ref{eq.varapprox}) indicates that coupling and mobility affect the variance differently. 
The first term in Eq.~(\ref{eq.varapprox}) is always negative and accounts for a decay in variance due to coupling.
%
%
The second and third terms are changes in variance due to mobility.
The second term is a mobility driven variance diffusion that can be positive or negative depending on variance concavity.
The third term is always positive, playing the role of a source of variance in the presence of a spatial phase gradient ${\bar \theta}(x,t)$ (Fig.~\ref{fig:S1}B). 
%
%
%
%
\begin{figure}[t]
\centering
\includegraphics[width=8.7 cm]{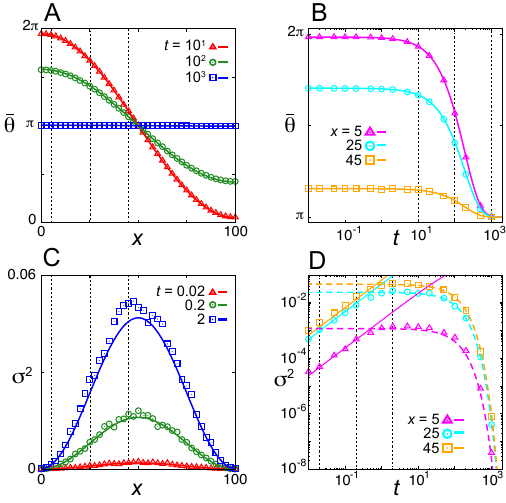}
\caption{Relaxation of the longest spatial mode in one-dimensional space.
(A), (B) Average phase ${\bar \theta}$ as a function of (A) position $x$ and (B) time $t$.
(C), (D) Phase variance $\sigma^2$ as a function of (C) position $x$ and (D) time $t$. 
Symbols indicate results of numerical simulations.
Colored lines in (A)-(C) indicate analytical solutions given by Eqs.~(\ref{eq.avephana1D}) and (\ref{eq.varana1D}).
Solid lines in (D) are solutions of $\partial \sigma^2/\partial t \approx {\bar \lambda} \left(\partial {\bar \theta} / \partial x\right)^2$.
Dashed lines in (D) show $\sigma^2 \approx ({\bar \lambda} /2\kappa) \left(\partial {\bar \theta} / \partial x\right)^2$.
Vertical dotted lines in (A),(C) indicate positions plotted in (B),(D), and those in (B),(D) indicate times plotted in (A),(C).
Parameters: $\kappa=1$, $\lambda / \kappa = 10$, $L = 100$, with $N = 1001$ and $\Delta x = 0.1$, and $n = 1$ in Eq.~(\ref{eq.fourier_initial}).
%
%
%
We use uniform coupling and mobility kernels with range $r_c = r_m = \sqrt{12}$.
Averages are computed over $10^3$ realizations.
}
\label{fig:1d_example}
\end{figure}

\subsection{Local fluctuations and the path to global synchrony}
We now look at solutions of the equations for the average phase and variance to clarify the relation between mobility induced local fluctuations and global synchronization.
We analyse the relaxation of spatial Fourier modes, since smooth spatial patterns can be represented as a superposition of such modes.
We consider the initial condition for the average phase and variance
\begin{equation}
{\bar \theta}(x, 0) = \pi \left( 1 + \cos\left( \frac{n \pi x}{L} \right) \right), \; \sigma^2(x, 0) = 0,
\label{eq.fourier_initial}
\end{equation}
where $n = 1, 2, 3, ...$ specifies a wave number.
%
%
The open boundary conditions for Eqs.~(\ref{eq.pde_ave}) and (\ref{eq.varapprox}) are $\partial_x {\bar \theta}(0, t) = \partial_x {\bar \theta}(L, t) = 0$ and $\partial_x \sigma^2(0,t) = \partial_x \sigma^2(L,t) = 0$.
Then, the solution of Eqs.~(\ref{eq.pde_ave}) and~(\ref{eq.varapprox}) is
\begin{equation}
{\bar \theta}(x, t) = \pi \left( 1 + e^{-\xi t} \cos\left( \frac{n \pi}{L}x \right) \right),
\label{eq.avephana1D}
\end{equation}
\begin{align}
\sigma^2(x, t) = \frac{n^2 \pi^4 {\bar \lambda} }{4 L^2} \left[ \frac{1}{\kappa-\xi} \left( e^{-2\xi t} - e^{-2\kappa t} \right) \right. \nonumber \\
\left.
- \frac{1}{\eta-\xi} \left( e^{-2\xi t} - e^{-2\eta t} \right) \, \cos\left( \frac{2 n\pi}{L}x \right) \right]
\label{eq.varana1D}
\end{align}
where
\begin{equation}
\xi = \frac{{\bar \kappa} + {\bar \lambda}}{2} \left(\frac{n \pi}{L} \right)^2, \; \eta = \kappa + {\bar \lambda} \left(\frac{n\pi}{L} \right)^2 \, .
\label{eq.xieta}
\end{equation}

We focus on the longest spatial mode $n = 1$, which often appears after transient in simulations of mobile oscillators as shown in Fig.~\ref{fig:x_phase}F.
The analytical results are in good agreement with numerical simulations for $\lambda/\kappa = 10$, Fig.~\ref{fig:1d_example}A-D.
%
%
Average phases approach their steady state value exponentially, with a rate $\xi$, Eqs.~(\ref{eq.avephana1D}) and (\ref{eq.xieta}).
In contrast, variance dynamics is characterized by the three timescales, $\eta^{-1}$, $\kappa^{-1}$ and $\xi^{-1}$.
For parameter values in Fig.~\ref{fig:1d_example} we obtain $\xi \approx 0.005 \kappa$ and $\eta \approx 1.01 \kappa$, so $\eta^{-1} \lesssim \kappa^{-1} \ll \xi^{-1}$.
This ordering is due to the timescale of mobility $({\bar \lambda}/L^2)^{-1} \gg 1$.
%
%
These timescales separate different dynamical regimes.
For very short times $t \ll \eta^{-1}$, we expand the exponentials in Eq.~(\ref{eq.varana1D}) and obtain a linear increase in the variance, resulting from the effect of mobility in the presence of a phase gradient $\partial_t \sigma^2 \approx {\bar \lambda} \left(\partial_x {\bar \theta}\right)^2$, solid lines in Fig.~\ref{fig:1d_example}D.
%
%
Thus, mobility induces local phase fluctuation at early times.
In contrast, for long times $t \gg \kappa^{-1}$ the exponential decrease $e^{-2\xi t}$ dominates the dynamics as the variance approaches zero.
From the solution, we infer that $\partial_t \sigma^2 \sim \partial_{xx} \sigma^2 \sim L^{-4}$, while $\sigma^2 \sim (\partial_x \bar\theta)^2 \sim L^{-2}$.
Neglecting the $L^{-4}$ terms in Eq.~(\ref{eq.varapprox}) we find that $\sigma^2 \approx ({\bar \lambda} /2\kappa) \left(\partial_x {\bar \theta} \right)^2$, dashed lines in Fig.~\ref{fig:1d_example}D.
This indicates that this regime is characterised by a balance between phase gradient variance production and variance decay due to coupling.
%
%
In this long time regime, mobility drives the variance decrease by decreasing phase gradients at a rate $\xi$, leading to synchronization.
%

\begin{figure}[t]
\centering
\includegraphics[width=8.3cm]{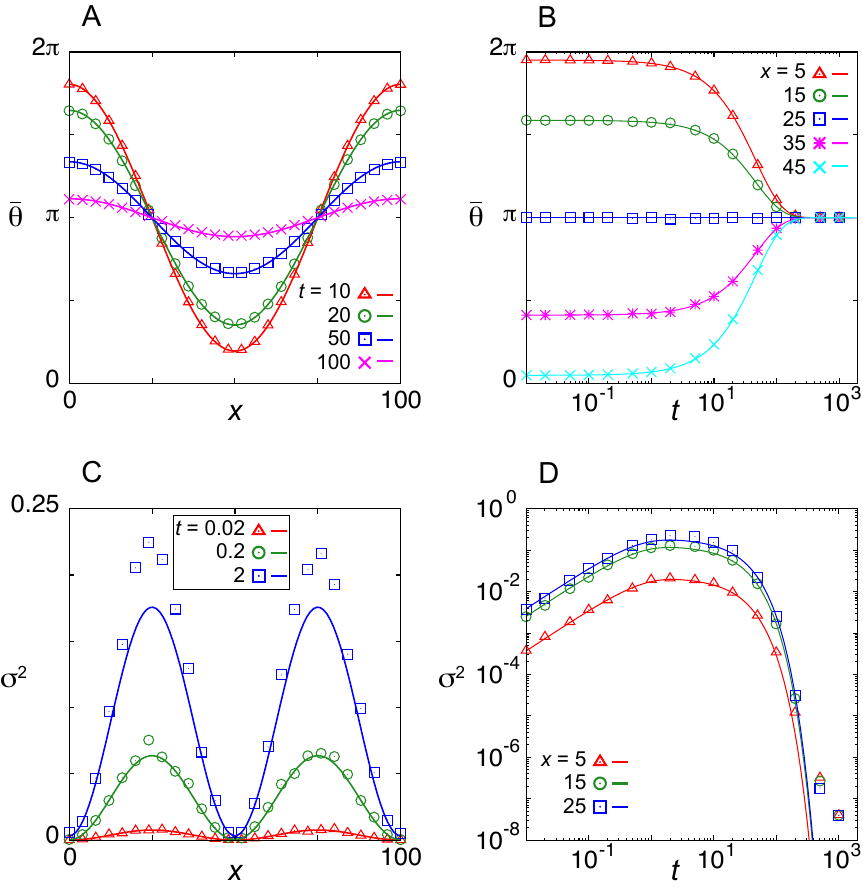}
\caption{Relaxation of a spatial Fourier mode with $n = 2$ in a one-dimensional lattice model.
(A, B) Average phase ${\bar \theta}(x,t)$ as a function of (A) $x$ for different time points, and (B) as a function of $t$ for different positions.
(C, D) Phase variance $\sigma^2(x, t)$ as a function of (C) $x$ for different time points, and (D) as a function of $t$ for different positions.
Symbols indicate results of numerical simulations. Lines indicate analytical solutions Eqs.~(\ref{eq.avephana1D})-(\ref{eq.xieta}).
Uniform coupling and mobility kernels are used with $r_c = r_m = \sqrt{12}$.
$\lambda/\kappa = 10$ with $\kappa = 1$. 
$\omega = 0$.
$L = 100$ with $\Delta x = 0.1$ and $N = 1001$.
%
%
}
\label{fig:2nd_mode}
\end{figure}

\begin{figure}[t]
\centering
\includegraphics[width=8.3cm]{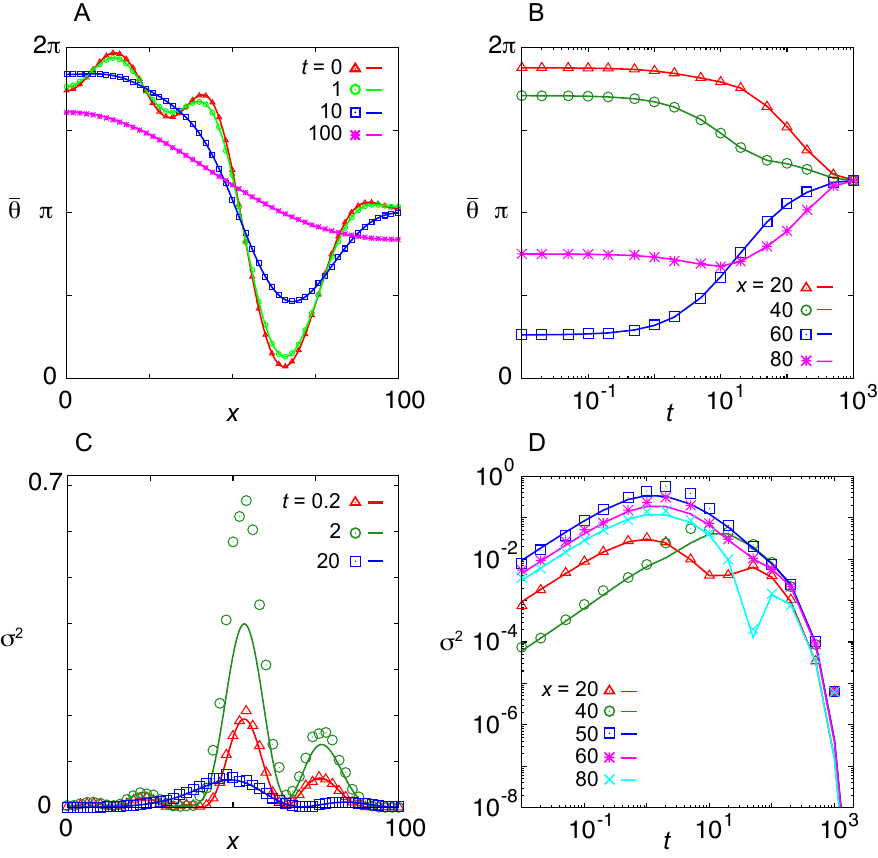}
\caption{Relaxation of a spatial phase pattern composed of mixed Fourier modes. 
%
%
(A, C) Spatial dependence of 
(A) average phase ${\bar \theta}(x,t)$ and 
(C) phase variance $\sigma^2(x,t)$ for different time points.
(B, D) Time evolution of 
(B) ${\bar \theta}(x,t)$ and 
(D) $\sigma^2(x,t)$ for different lattice sites.
Symbols indicate results of $10^3$ realizations. 
Lines indicate numerical solutions for average phase Eq.~(\ref{eq.pde_ave}) and variance Eq.~(\ref{eq.varapprox}). 
The uniform coupling and mobility kernels are used in the calculations, $r_c = r_m = \sqrt{12}$.
$\lambda/\kappa = 10$ with $\kappa = 1$. 
$\omega = 0$.
The number of oscillators is $N = 1001$, separated by $\Delta x = 0.1$.
%
Initial condition is the same as in Fig.~\ref{fig:x_phase}.
%
}
\label{fig:mixed_modes}
\end{figure}

\begin{figure}[tb]
\centering
\includegraphics[width=8.3 cm]{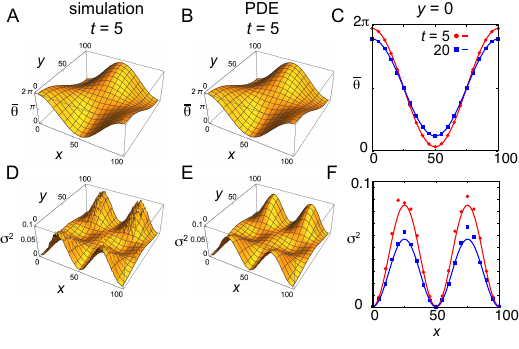}
\caption{Relaxation of a spatial mode in two-dimensional lattice.
(A, B) Average phase ${\bar \theta}(x,t)$ in a 2D lattice ($100 \times 100$) obtained by 
(A) numerical simulation of Eq.~(\ref{eq.phase_disc_2D}) and 
(B) the PDE Eq.~(\ref{eq.pde_ave_2D}) in Appendix F. 
(C) Average phase in a y-slice at $y = 0$.
(D, E) Phase variance $\sigma^2(x,t)$ obtained by 
(D) numerical simulation of Eq.~(\ref{eq.phase_disc_2D}) and 
(E) PDE Eq.~(\ref{eq.var_dot_approx_2D}).
(F) Phase variance in a y-slice at $y = 0$.
In (C) and (F), symbols indicate results of direct simulations and colored lines indicate the solutions of PDE.
$\lambda / \kappa = 10$ with $\kappa = 1$. 
${\bar \theta}$ and $\sigma^2$ for direct simulations are calculated from $10^3$ realizations.
We used an isotropic lattice space $\Delta x = \Delta y =  1$.
The initial condition is 
$\theta(x, y, 0) = \pi \left( 1 + \cos \left( {n_x \pi x/L} \right) \cos \left( {n_y \pi y /L} \right) \right)$, 
with $n_x = 2$ and $n_y = 1$.
}
\label{fig:2d_example}
\end{figure}

We confirmed similar behaviors for shorter spatial modes (Fig.~\ref{fig:2nd_mode}).
Relaxation of a shorter mode ($n = 2$) occurs faster than that of the longest mode ($n = 1$) as predicted by Eqs.~(\ref{eq.avephana1D})-(\ref{eq.xieta}), (Fig.~\ref{fig:2nd_mode}A, B). 
%
%
While the solution of PDE for average phase consistently provides a good approximation of numerical simulations, the approximation for the variance becomes less precise at longer times for shorter modes (Fig.~\ref{fig:2nd_mode}C, D).
%
%
%
%
%
%
%
Although these deviations in the variance set some limitations, we confirm overall agreements between the analytical calculation and simulations.
We also confirmed that the theory provides a good description of the relaxation of mixed spatial modes (Fig.~\ref{fig:mixed_modes}) and in 2D (Fig.~\ref{fig:2d_example}, Appendix F).
\begin{figure*}[t]
\centering
\includegraphics[width=15cm]{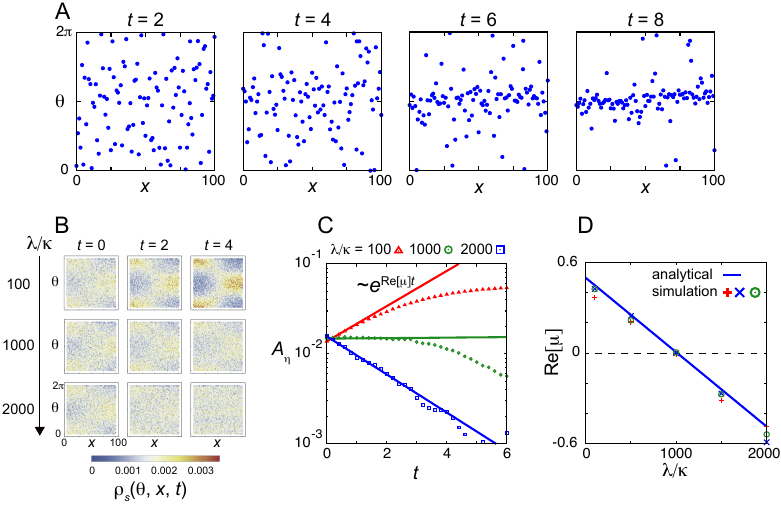}
\caption{Mean-field transition in a one-dimensional lattice model. 
(A) Spatial phase patterns at $t = 2, 4, 6$ and $8$ for $\lambda/\kappa = 1050$. Each blue dot indicates phase value of an oscillator.
Simulations are started with random initial phases in the interval $[0, 2\pi)$.
(B) Time evolution of numerically constructed probability density $\rho_s(\theta, x, t)$ in the one-dimensional lattice model.
(C) Amplitude $A_{\eta}(t)$ of the longest spatial mode in $(\eta_{1}(x,t)+\eta_{-1}(x,t))/2$ for different values of $\lambda/\kappa$ (symbols).
Lines indicate $A_{\eta}(0)e^{{\rm Re}[\mu]t}$ where $\mu$ is described in Eq.~(\ref{eq.mu}) in Appendix G.
(D) Exponential growth rate ${\rm Re}[\mu]$ of $A_{\eta}$ as a function of $\lambda/\kappa$.
Blue line indicates analytical result in Eq.~(\ref{eq.mu}).
Three different symbols indicate numerically obtained values from three independently constructed $\rho_s$.
To calculate $\rho_s$ and $A_{\eta}$, we performed $10^3$ realizations of simulations of mobile oscillators with each $\lambda/\kappa$. See Appendix A for computation of $A_{\eta}$.
$\kappa = 1$ and $\omega = 0$.
$N = 101$ with $\Delta x = 1$.
}
\label{fig:mf}
\end{figure*}

\twocolumngrid

\subsection{Mean-field transition}
%
%
For fast mobility $\lambda / \kappa \gg 1$, local phase fluctuations become large, preventing the formation of spatial modes (Fig.~\ref{fig:mf}A and Appendix A) ~\cite{uriu13, petrungaro19}.
%
%
%
Thus, there might be a transition point where spatial modes cannot grow from random initial phases and the path to synchronization qualitatively changes.

For fast mobility $\lambda / \kappa \gg 1$, the correlation of phases between neighboring lattice sites would be small, 
so we can approximate the joint probability density as $\rho(\theta', x', t; \theta, x, t) \approx \rho(\theta', x', t) \rho(\theta, x, t)$ in Eq.~(\ref{eq.master_preexpan}).
%
%
We express the $2 \pi$-periodic function $\rho(\theta, x, t)$ for $\theta$ using Fourier series,
\begin{equation}
\rho(\theta, x, t) = \frac{1}{2 \pi} + \sum_{l \neq 0} \eta_l (x, t) e^{i l \theta},
\label{eq.rho_fourier}
\end{equation}
with the normalization condition $\int_{0}^{2 \pi} d\theta \rho(\theta, x, t) = 1$~\cite{strogatz91}.
%
%
Substituting Eq.~(\ref{eq.rho_fourier}) into Eq.~(\ref{eq.master_preexpan}) and collecting terms including $e^{il\theta}$, we obtain a system of coupled ODEs for the Fourier coefficients $\eta_l(x,t)$.
Equations for $\eta_{\pm 1}(x,t)$ include linear growth terms $\sim \eta_{\pm 1}(x,t)$, whereas those for other $\eta_l(x,t)$ ($l \neq \pm 1$) consist of higher order terms (Appendix G).
%
%
Hence, we expect $\eta_{\pm 1}$ to increase first, followed by other $\eta_l$'s.
%
%
Synchronization dynamics qualitatively changes when spatially heterogeneous components of $\eta_{\pm 1}(x,t)$ do not grow due to fast mobility. 
This predicts the transition point at (Appendix G):
\begin{equation}
\frac{{\bar \lambda}}{\kappa} \approx \left( \frac{L}{\pi} \right)^2 -\frac{C_2}{2} \approx \left( \frac{L}{\pi} \right)^2.
\label{eq.mfonset}
\end{equation}
To confirm this prediction, we construct the numerical probability density from simulations (Fig.~\ref{fig:mf}B), and calculate $\eta_{\pm 1}(x,t)$ by Fourier series analysis.
From these we obtain the amplitude of the longest spatial modes $A_{\eta}(t) = (1/L)\int^{L}_{0}dx (\eta_{1}(x,t) +\eta_{-1}(x,t) ) \cos(\pi x /L)$ (Appendix A).
As expected, these amplitudes grow for slow mobility and decay for fast mobility (Fig.~\ref{fig:mf}C).
Growth rates obtained from simulations are in very good agreement with the analytical expressions from the statistical description (Fig.~\ref{fig:mf}D, Appendix G).
As predicted, amplitude growth becomes vanishingly small close to the predicted value of mobility, $\lambda/\kappa \approx 1033$ (Fig.~\ref{fig:mf}C, D).
Previous studies indicated that if $\lambda / \kappa \sim L^2$, a population of mobile oscillators with nearest neighbor coupling in a one-dimensional lattice behaves as they are coupled to the mean-field (i.e., all-to-all coupling)~\cite{uriu13,petrungaro19}.
Thus, the transition point Eq.~(\ref{eq.mfonset}) would mark the onset of such mean-field behavior.
%


\section{Statistical description in inhomogeneous space}
\begin{figure}[t!]
\centering
\includegraphics[width=8.5cm]{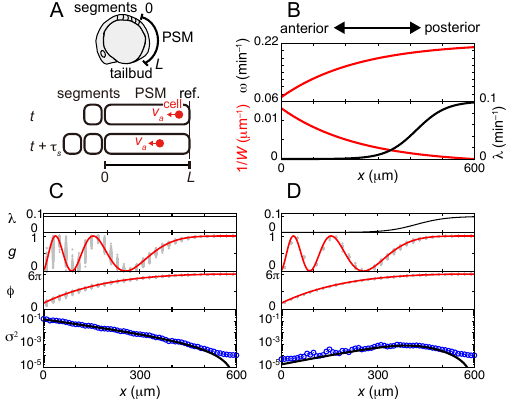}
\caption{Robustness of wave patterns in the presence of a cell mobility gradient.
(A) Schematic of zebrafish segmenting tissue.
In the reference frame (ref.) at the tail of an embryo, cells advect anteriorly due to tissue elongation at speed $v_a$.
$\tau_s$ is the segmentation period.
(B) Spatial gradients of autonomous frequency $\omega$, inverse of wavelength $1/W$ and mobility rate $\lambda$.
(C), (D) Spatial dependence of gene expression $g = (1+\cos\phi)/2$, phase $\phi$ and phase variance $\sigma^2$ with (C) uniform mobility and (D) mobility gradient.
To obtain the steady state phase pattern, we first estimate the collective frequency $\Omega_s$ from a linear fit of the phase $\theta(t)$ at the posterior after transient, and then subtract $\Omega_s t$ from phase values.
Gray dots are gene expression and phase from $10^{3}$ numerical simulations of Eq.~(\ref{eq.fgad_orig}).
Blue circles indicate phase variance from the same data. 
Red lines show gene expression and phase from approximation Eq.~(\ref{eq.phi_approx}). 
Black lines indicate the approximate variance solution Eq.~(\ref{eq.sigma_approx2}).
Initial condition is $\theta_i(0) = 6 \pi$ for all $i$.
Parameters: $v_a = 0.167$, $\kappa = 0.07$, $\omega_L = 2\pi/30$, $\nu = 0.34$, $k = 2.5$, $N = 61$, $L=600$, $\beta=12.8$, and $\lambda_L=0.1$. $x_h =-3.0$ in (C) and $x_h = 0.7$ in (D).
}
\label{fig:wave_mobility}
\end{figure}

%
%
In previous sections we studied how synchronization is affected by mobility in the posterior of the segmenting tissue, assuming homogeneous space.
Next, we extend the statistical description to investigate pattern robustness across the entire segmenting tissue (Fig.~\ref{fig:wave_mobility}A).
%
%
%
As described in the introduction, the segmenting PSM encompasses oscillator coupling, with a frequency gradient that causes traveling phase waves, and a cell mobility gradient.
Additionally, tissue elongation causes advective cell motion relative to an observer at the tail of the embryo (Fig.~\ref{fig:wave_mobility}A)~\cite{jorg15}.

Motivated by the finite cell size and cell's volume exclusion, we consider a one-dimensional lattice describing the tissue along the anterior-posterior axis of embryos (Fig.~\ref{fig:wave_mobility}A).
As a newly formed segment buds off from the anterior unsegmented tissue, tissue length becomes shorter.
For simplicity, here we consider a constant tissue length $L$, assuming that continuous elongation at the posterior compensates this anterior shortening.
With a lattice distance $\Delta x$ representing the cell diameter, the total number of lattice sites is $N = L / \Delta x$.
We set $x = 0$ at the anterior end and $x = L$ at the posterior tip of the tissue, so the reference frame moves with the tissue domain (Fig.~\ref{fig:wave_mobility}A)~\cite{jorg15}.

Coupled phase oscillators in this lattice are described by~\cite{morelli09}
\begin{align}
\frac{d \theta_i(t)}{dt} = \omega_i &+ v_a (\theta_{i+1}(t) - \theta_{i}(t)) \nonumber \\
&+ \frac{\kappa}{n_i} \sum_{j = i \pm 1} \sin \left(\theta_j(t) - \theta_i(t) \right) \, ,
\label{eq.fgad_orig}
\end{align}
where $\theta_i(t)$ is the phase at site $i$ and time $t$.
The spatial dependence of autonomous frequency $\omega_i$ is determined from experimental observations (Fig.~\ref{fig:wave_mobility}B; Appendix A)~\cite{jorg15,soroldoni14}.
The second term in Eq.~(\ref{eq.fgad_orig}) represents the phase advection~\cite{morelli09}.
We consider uniform advection velocity $v_a \Delta x$, since the elongating zone is primarily restricted to the posterior tip of the tissue~\cite{mongera18}.
The third term represents coupling between oscillators at neighboring lattice sites.
Cell mobility is larger in the posterior region of the tissue, decreasing towards the anterior~\cite{lawton13, uriu17b, mongera18}.
We model the mobility gradient with a sigmoidal function (Fig.~\ref{fig:wave_mobility}B, Appendix A).

%
Since experimental observation of patterns revealed a homogeneous region at the posterior~\cite{soroldoni14}, we assume the boundary condition $\theta_{N+1}(t)=\theta_{N}(t)$. 
Beyond the anterior boundary the oscillations stop and the pattern is advected, so $\omega_i = \kappa_i = 0$ for all $i<0$.
%
%
For numerical simulations, we use the time and length scales of segment formation in zebrafish embryos~\cite{uriu17b, jorg15, riedel07}.
Typical cell diameter is $\sim 10 \ \mu {\rm m}$~\cite{uriu17b}, which suggests $\Delta x = 10 ~\mu {\rm m}$. 
The period of segmentation in zebrafish is around 30 min~\cite{oates12}, which suggests $\omega(L) = \omega_L = 2\pi/30$ min$^{-1}$. 
Previous studies estimated $\kappa \approx 0.1$ min$^{-1}$~\cite{riedel07, herrgen10} and $\lambda(L) \approx 0.05 \sim 0.1$ min$^{-1}$~\cite{uriu17b}.
%

To illustrate how mobility affects phase variance in the presence of persistent phase patterns, we first consider the simpler case of uniform mobility.
We find that mobility enhances phase variance production in the presence of the anterior wave pattern, in contrast to the posterior (Fig.~\ref{fig:wave_mobility}C).
In the presence of a mobility gradient as observed in embryos, phase variance is very small both at anterior and posterior regions (Fig.~\ref{fig:wave_mobility}D).
%

To clarify the relation between persistent spatial patterns and phase variance, we apply the statistical description.
We derive the time evolution of average phase (Appendix H),
\begin{align}
\frac{\partial  \bar\theta(x,t) }{\partial t}  = \omega(x) + {\bar v}_a \frac{\partial \bar\theta(x,t)}{\partial x} 
+  \frac{{\bar \kappa}  + {\bar \lambda}(x)}{2} \frac{\partial^2  \bar\theta(x, t)}{\partial x^2},
\label{eq.phi}
\end{align}
%
%
with the scaling, $v_a = {\bar v}_a/ \Delta x$, $\kappa = {\bar \kappa}/ \Delta x^2$ and $\lambda(x) = {\bar \lambda}(x)/ \Delta x^2$.

We introduce the steady state ansatz $\bar\theta(x,t)=\phi(x)+\Omega t$~\cite{morelli09}, and the posterior boundary condition $d\phi / dx |_{x=L} = 0$, together with the approximation $d^2\phi / dx^2 |_{x=L} = 0$ to describe flat posterior pattern.
We obtain the collective frequency $\Omega = \omega(L)$ and the phase profile ODE
\begin{equation} \label{eq.phiode}
\frac{\bar\kappa+\bar\lambda(x)}{2} \phi''(x) + \bar{v}_a \phi'(x) + \omega(x) - \omega(L) = 0 \,.
\end{equation}
Assuming that the effects of coupling and mobility on phase gradient are smaller than those of frequency and advection~\cite{jorg15}, we neglect the second spatial derivative terms in Eq.~(\ref{eq.phiode}) 
\begin{equation}
\phi'(x) \approx \frac{\omega(L)-\omega(x)}{\bar v_a} \,,
\label{eq.dphi_approx}
\end{equation}
and solve for the phase
\begin{equation}
\phi(x) \approx \phi(0) + \int_{0}^{x} dx' \frac{\omega(L) - \omega(x')}{{\bar v}_a},
\label{eq.phi_approx}
\end{equation}
see Fig.~\ref{fig:wave_mobility}C, D.
%
%
%
%

For the variance $\sigma^2(x, t)$, we follow the same approach of previous sections with truncation of covariance terms (Appendix H) and obtain
\begin{align}
\frac{\partial \sigma^2 (x,t)}{\partial t} \approx - 2 \left(\frac{{\bar \kappa}}{\Delta x^2} + \frac{{\bar v}_a}{\Delta x} \right) \sigma^2(x,t) + \frac{{\bar \lambda}(x)}{2} \frac{\partial^2 \sigma^2 (x,t)}{\partial x^2} \nonumber \\ +  {\bar \lambda}(x) \left( \frac{\partial \bar\theta (x,t)}{\partial x} \right)^2.
\label{eq.var_fgncv}
\end{align}
The first term in Eq.~(\ref{eq.var_fgncv}) indicates that variance is reduced both by coupling and advection.
%
%
The second and third terms represent variance diffusion and production caused by cell mobility.
To find an approximate solution for the variance, we make the steady state ansatz for $\bar\theta(x,t)$ together with $\partial_t \sigma^2(x,t)=0$. 
Neglecting variance diffusion we arrive at 
\begin{align}
\sigma^2(x) \approx \frac{{\bar\lambda}(x) \Delta x^2}{2 ({\bar \kappa} + {\bar v}_a \Delta x)}  \left( \frac{d\phi}{dx} \right)^2.
\label{eq.sigma_dphi}
\end{align}
This shows how phase variance results from the presence of mobility ${\bar\lambda}$ and a steady state phase gradient $d\phi/dx$.
The spatial gradient of $\phi$ is larger for smaller $x$ due to the imposed frequency gradient $\omega (x)$ (Fig.~\ref{fig:wave_mobility}B-D).
As a result, the variance $\sigma^2$ becomes large toward smaller $x$ (Fig.~\ref{fig:wave_mobility}C).
We can replace the phase gradient Eq.~(\ref{eq.dphi_approx}) to obtain
\begin{equation}
\sigma^2(x) \approx \frac{{\bar \lambda}(x) \Delta x^2}{2 ({\bar \kappa} + {\bar v}_a \Delta x)} \left(  \frac{\omega(L) - \omega(x)}{{\bar v}_a} \right)^2.
\label{eq.sigma_approx2}
\end{equation}
This expression reveals that variance is related to the difference between the local frequency and collective frequency $\Omega=\omega(L)$.
This result is in very good agreement with numerical simulations, Fig.~\ref{fig:wave_mobility}C, D.

Introducing the local wavelength of the spatial pattern $W(x) \approx 2 \pi / |d\phi(x)/dx|$~\cite{jorg15, giudicelli07}, the variance at the steady state can be written as
\begin{align}
\sigma^2(x) \approx \frac{{\bar \lambda}(x)}{2 ({\bar \kappa} + {\bar v}_a \Delta x)}  \left( \frac{2 \pi \Delta x}{W(x)} \right)^2.
\label{eq.sigma_approx}
\end{align}
This equation also reveals the relation between mobility and pattern wavelength.
A phase pattern is less sensitive to mobility if pattern wavelength is long, 
$W/{\Delta x} \gg 1$.
%
In contrast, a phase pattern with a short wavelength is sensitive to mobility.
In the zebrafish PSM, the wavelength of gene expression pattern becomes shorter in the anterior region (Fig.~\ref{fig:wave_mobility}B).
The opposing gradients of wavelength and cell mobility reduce phase variance across the zebrafish PSM (Fig.~\ref{fig:wave_mobility}B, D).
Hence, the gradient of cell mobility across the PSM is important to maintain integrity of gene expression waves at the anterior part of the tissue, while to enhance synchronization at its posterior part.
%

\section{Discussion}
In this study, we derived a statistical description of mobile phase oscillators.
Various techniques have been applied to analyze synchronization of mobile oscillators, including master stability function combined with fast switching approximation~\cite{skufca04,frasca08}, spectral analysis of time varying Laplacian matrix~\cite{fujiwara11}, and hydrodynamic equations~\cite{grossmann16,levis19}.
Hydrodynamic theory uses probability density for mobile particles and coarse-graining by the calculation of a local order field~\cite{grossmann16,levis19}, which would be the counterparts of Eqs.~(\ref{eq.master_preexpan}) and~(\ref{eq.pde_ave}).

The statistical description derived here reveals the synchronization mechanism of mobile oscillators by contrasting the dynamics of average phase and variance.
On one hand, neighbor exchange causes local fluctuations in the presence of a phase gradient.
On the other hand, neighbor exchange reduces the phase gradient together with oscillator coupling.
Mobility driven variance production dominates an initial regime until this is balanced by coupling.
After this regime, mobility driven gradient decrease dominates and variance is reduced at a faster rate.
%
%
Hence, mobility promotes synchronization across space.
For fast mobility, fluctuation dominates and no spatial pattern forms.
We identify the scaling between mobility, coupling and system size that sets the onset of the longest spatial mode instability, revealing the onset of mean-field behavior.
Thus, the statistical description reveals the relation between fluctuation and synchronization.
Several previous studies considered complex movement patterns, such as collective motions~\cite{uriu14a}, super diffusion~\cite{grossmann16}, and oscillators phase dependent swarming~\cite{tanaka07, okeeffe17, levis19,okeeffe22,sar22}.
Deriving a statistical description with such complex movement patterns would be an interesting future work.

Pattern formation under cell mobility is a hallmark of development.
In some cases, cell movement itself generates periodic spatial patterns with a characteristic length scale as in zebrafish skin strip formation~\cite{nakamasu09}.
In some other cases, cell movement may challenge gene expression patterning, blurring boundaries through mixing~\cite{fulton22}.
%
The statistical description revealed a generic relationship between mobility and pattern wavelength for the robustness of spatial patterns.
In long wavelength patterns, mobility promotes synchronization without increasing variance.
In contrast, in short wavelength patterns mobility increases phase variance due to steep phase gradients.
%
%
%
Recent imaging of mRNA and protein cycles at single cell level in the embryo~\cite{eck24, rohde24} may open the road to test these predictions.

In the embryo, the mobility gradient is generated by signaling gradients across the tissue.
The concentration of fibroblast growth factor (FGF) is higher at posterior region than anterior region in the PSM, and FGF signal is known to promote cell mobility~\cite{benazeraf10, delfini05, lawton13}.
Recent studies also pointed out an involvement of a jamming transition for regulating cell mobility~\cite{mongera18}.
Such state transition is necessary to drive axis elongation of embryos~\cite{mongera18, banavar21}.
Here, we demonstrate that a change in mobility may be also important for correct segment boundary formation.

The statistical description we introduced here is of broad applicability. 
The resulting relation between variance and pattern wavelength is general, independent of the particular form of frequency and mobility spatial profiles.
Additionally, while here we considered mobile coupled oscillators, the general framework we presented in this work may be useful to tackle other systems with collective patterning by mobile agents or elements.
We expect that this approach may provide insight on the relation between mobility-induced fluctuations and pattern formation more generally.

\subsection*{ACKNOWLEDGMENTS}
We thank Sa{\'u}l Ares, Gabriela Petrungaro and Olivier Venzin for their helpful comments on the manuscript.
This work was partly supported by the JSPS KAKENHI grant number 19H04772 and 20K06653 to KU, ANPCyT grant PICT 2019 0445 awarded to LGM and FOCEM-Mercosur (COF 03/11) to IBioBA. 
We thank JSPS Long Term Invitational Fellowship L23529, and LGM thanks the Uriu Lab and Tokyo Tech for hospitality.
LGM is a researcher of CONICET.


\appendix

\section{Methods}
\subsection{Coupling and mobility kernels}
We use uniform coupling and mobility kernels 
\begin{equation} \label{eq.kernels}
C(x) = \frac{1}{r_c} \, \Theta\left(\frac{r_c}{2} - |x| \right)
, \quad
M(x) = \frac{1}{r_m} \, \Theta\left(\frac{r_m}{2} - |x| \right) \, ,
\end{equation}
where $\Theta$ is the Heaviside step function.
In the lattice model Eq.~(\ref{eq.phase_cont}), $x_i = i \Delta x$, $C_{ij} = C(x_j - x_i) \Delta x$ and 
$M_{ij} = M(x_j - x_i) \Delta x$.
We employ open boundary conditions for both coupling and mobility (Fig.~\ref{fig:x_phase}B).
%
%
An oscillator near the left boundary $0 \leq x < r_m/2$ may exchange its position with an oscillator in the interval $[0, x+r_m/2]$ with probability $(x+r_m/2)^{-1} \Delta x$.
%
%
Similarly, an oscillator near the right boundary $x > L-r_m/2$ may exchange its position with an oscillator in the interval $[x-r_m/2,L]$ with probability $(L-x+r_m/2)^{-1} \Delta x$.
%
%
These boundary conditions of the mobility kernel keep the normalization $\int^{L}_{0} dx' M(x'-x) = 1$, but introduce differences in its first and second moments between oscillators in bulk and those near boundaries.
%
%
%
%
In general, the $n$th moment of a normalized kernel $G(x'-x)$ is defined as
\begin{equation}
    G_n(x) \equiv \int_{0}^{L} dx' (x'-x)^n G(x'-x).
\end{equation}
If the kernel is an even function $G(x) = G(-x)$ and its spatial range $r_G$ where $G(x) \neq 0$ is  restricted $r_G/L \ll 1$, then $G_{2m+1}(x) = 0$ in the bulk space defined as $r_G/2 \leq x \leq L-r_G/2$.
In contrast, $G_{2m}$ is constant in the bulk, especially $G_0 = 1$ because of the normalization.
With $r_c = r_m = \sqrt{12}$, the first and second moments in the bulk are $C_1 = M_1 = 0$ and $C_2 = M_2 = 1$, respectively.

%
\begin{table}[t]
 \begin{center}
   \caption{Integration timesteps for numerical simulations}
   \label{table:ts}
  \begin{tabular}{lll} \hline
    $\lambda / \kappa$ & $dt$ & Figure \\ \hline
    0 & $10^{-2}$ & Fig.~\ref{fig:x_phase} \\ 
    $10^{-1}$ & $2 \times 10^{-3}$ & Fig.~\ref{fig:x_phase} \\
    $10^{0}$ & $5 \times 10^{-4}$ & Fig.~\ref{fig:x_phase} \\
    $10^{1}$ (1D lattice) & $10^{-4}$ & Figs.~\ref{fig:x_phase}-\ref{fig:mixed_modes} \\
    $10^{1}$ (2D lattice) & $10^{-5}$ & Fig.~\ref{fig:2d_example} \\
    $10^{2}$ & $10^{-5}$ & Fig.~\ref{fig:mf} \\
    $5 \times 10^{2}$ & $10^{-5}$ & Fig.~\ref{fig:mf}\\
    $10^{3}$ & $5 \times 10^{-6}$ & Fig.~\ref{fig:mf} \\
    $1.5 \times 10^{3}$ & $2 \times 10^{-6}$ & Fig.~\ref{fig:mf}\\
    $2 \times 10^{3}$ & $2 \times 10^{-6}$ & Fig.~\ref{fig:mf}\\
    $1.43$ ($\lambda_L / \kappa$) & $10^{-3}$ & Fig.~\ref{fig:wave_mobility} \\ \hline
  \end{tabular}
 \end{center}
\end{table}

\subsection{Initial conditions}
The initial condition used in Fig.~\ref{fig:x_phase} is 
%
%
$\theta(x_i, 0) = \sum^{9}_{j=0} a_j \cos\left( \pi j x_i/L \right)$,
with $a_0 = 3.76$, $a_1 = 2.10$, $a_2 = 0.757$, $a_3 = -1.26$, $a_4 = 0.393$, $a_5 = 0.424$, $a_6 =-0.561$, $a_7 = -0.201$, $a_8 = -0.00318$ and $a_9 = 0.0454$.

\subsection{Fourier modes near the mean-field transition}
To confirm the analytical calculation for mean-field transition, we numerically construct probability density $\rho(\theta, x, t)$ and analyze the initial growth of Fourier modes.
In a lattice model with nearest neighbor exchanges, Eq.~(\ref{eq.mfonset}) can be represented as $\lambda/\kappa \approx (N/\pi)^2$ because ${\bar \lambda}=\lambda \Delta x^2$ and $L = N \Delta x$. 
The initial phase of each oscillator is randomly chosen following the probability density $\rho(\theta, x, 0)=1/(2 \pi) + \epsilon \cos(\pi x/L) \cos\theta$ with $\epsilon = 0.03$.
$10^3$ simulations are performed and $\rho$ at each time point is obtained from the ensemble simulation data with the bin size of $\Delta \theta = 0.3$ for phase and $\Delta x = 1$ for lattice space (Fig.~\ref{fig:mf}B).
We then numerically calculate ${\bar \eta}_1(x, t) \equiv (\eta_1 + \eta_{-1})/2 = (2\pi)^{-1} \int^{2 \pi}_{0} d\theta \rho(\theta, x, t) \cos\theta$.
Subsequently, we obtain the amplitude $A_\eta(t)$ of the longest spatial mode of ${\bar \eta}_1$, as defined in the main text.
%
%
The growth rate of $A_\eta(t)$ is estimated by fitting an exponential function $a_s e^{\mu t}$ for the numerical data of $A_\eta(t)$ for $0 \leq t \leq 1$.
\subsection{Frequency and mobility gradients in body segment formation}
Based on previous studies~\cite{jorg15, soroldoni14}, we model the frequency gradient in zebrafish embryos as
\begin{equation}
\omega(x) = \omega_L \left( \nu + (1-\nu) \frac{1-e^{-kx/L}}{1-e^{-k}} \right),
\label{eq.omega}
\end{equation}
where $\omega_L$ is the maximum frequency at $x = L$, $\nu$ is the ratio $\omega(0)/\omega(L)$, and $k$ is the shape parameter of the frequency gradient.
$\omega(x)$ in Eq.~(\ref{eq.omega}) provides a monotonic gradient where the autonomous frequency increases with $x$.
For the cell mobility gradient, we assume
\begin{equation}
\lambda(x) = \frac{\lambda_L}{1 + \exp(-\beta (x - x_h))},
\label{eq.mobility_grad}
\end{equation}
where $\lambda_L$ is the mobility rate at the posterior tip of the PSM, $\beta$ is the steepness of the gradient and $x_h$ is the position at which mobility rate becomes half maximum $\lambda(x_h) = \lambda_L/2$.
Setting $x_h \to -\infty$ models spatially uniform mobility rate.
In the lattice model Eq.~(\ref{eq.fgad_orig}), $x_i = i \Delta x$, $\omega_i = \omega(x_i)$, and $\lambda_i = \lambda(x_i)$.
%

\subsection{Numerical simulations}
The timing of phase exchange events is modeled as a homogeneous Poisson process with rate $\lambda(x)$. 
%
Two locations that exchange phase values are chosen based on the mobility kernel $M_{ij}$ in Eq.~(\ref{eq.kernels}).
Between two successive exchange events, Eq.~(\ref{eq.phase_cont}) is numerically integrated by the second order Runge-Kutta method.
The time step $dt$ for the Runge-Kutta method was fixed during a single realization of simulations.  
Its value was adapted to the mobility rate so that the number of exchange events in the entire lattice within $dt$ was smaller than 3, see Table~\ref{table:ts} for values.
%
%
To obtain numerical solutions of PDEs for average phase and variance we used the explicit Euler method, with central spatial difference scheme. 
The custom codes for simulations are written in C language. 

\onecolumngrid

\section{Derivation of probability density dynamics Eq.~(2)}
Here we derive the time evolution of probability density $\rho(\theta, x, t)$ for mobile oscillators. 
We begin with Eq.~(1) of the main text for mobile oscillators in a discrete lattice, 
\begin{equation} \label{eq.phase_disc}
\dot\theta_i(t) = \omega + \kappa \sum_{j=0}^{N} {C_{ij} \sin\left(\theta_j(t) - \theta_i(t)\right)} \, .
\end{equation}
For small lattice spacing $\Delta x /L \ll 1$, we can take the limit $\Delta x \to 0$ and write the sum as an integral, 
%
%
\begin{equation}
\frac{\partial \theta(x, t)}{\partial t} = \omega + \kappa \int_0^L dx' C(x'-x) \sin \left( \theta(x',t) - \theta(x,t) \right) \,,
\label{eq.phase_cont2}
\end{equation}
where $C(x'-x)$ is the even, normalized coupling kernel, and spatial coordinate $x$ is now a continuum variable.
%
%
%

We consider a large ensemble of independent systems, and introduce $\rho(\theta, x, t)$ describing the probability density to observe a phase $\theta$ at position $x$ and time $t$.
%
%
The change in probability density in a small time interval $\Delta t$ is
%
%
\begin{equation}
\rho(\theta, x, t+\Delta t) - \rho(\theta, x, t) = \int_0^{2\pi} d\theta' P\left(\theta, x, t+\Delta t | \theta', x, t \right)\rho(\theta', x, t) - \int_0^{2\pi} d\theta' P\left(\theta', x, t+\Delta t | \theta, x, t \right)\rho(\theta, x, t),
\label{eq.master}
\end{equation}
%
%
%
%
%
where the first and second terms of Eq.~(\ref{eq.master}) represent gain and loss of probability, respectively.
%
%
The conditional probability $P(\theta_1,x,t+\Delta t | \theta_0,x,t)$ to observe phase $\theta_1$ at position $x$ and time $t+\Delta t$ given that it was $\theta_0$ at $x$ and time $t$, includes the deterministic dynamics of coupled phase oscillators and the stochastic dynamics of phase exchange,
\begin{equation}
P(\theta_1,x,t+\Delta t | \theta_0,x,t) = (1 - \lambda \Delta t) \delta(\theta_1 - \left\{\theta_0+F(\theta_0,x, t)\Delta t\right\}) + \lambda \Delta t\int_0^L dx' M(x'-x) \rho(\theta_1, x', t | \theta_0, x, t),
\label{eq.transition_gain}
\end{equation}
%
%
%
%
%
where $\delta$ is the Dirac delta function,
%
%
%
%
and we introduced the deterministic phase velocity
\begin{equation}
F\left(\theta, x,t\right) = \omega + \kappa \int_0^{2\pi} d\theta' \int_0^L dx' C(x'-x) \sin \left(\theta' - \theta \right) \rho(\theta', x', t | \theta, x, t).
\label{eq.F}
\end{equation}
%
%
%
%

Next, we substitute the conditional probability Eq.~(\ref{eq.transition_gain}) into the gain and loss terms in Eq.~(\ref{eq.master}).
We first consider the gain term,
\begin{align}
\int_0^{2\pi} d\theta' P\left(\theta, x, t+\Delta t | \theta', x, t \right)\rho(\theta', x, t) = 
(1-\lambda \Delta t) & \int_{0}^{2\pi} d\theta' \delta \left( \theta - (\theta'+F(\theta',x,t) \Delta t) \right) \rho (\theta',x,t) \nonumber \\
& + \lambda \Delta t \int_{0}^{2\pi} d\theta' \int_0^L dx' M(x'-x) \rho(\theta, x', t | \theta', x, t) \rho(\theta',x,t) \, .
\end{align}
For the small time interval $\Delta t \ll 1$ we can approximate $\theta' - \theta \equiv \varepsilon \ll 1$.
This allows expansion of $F(\theta',x,t)$ in the argument of the Dirac delta,
\begin{equation}
\int_{0}^{2\pi} d\theta' \delta \left( \theta - (\theta'+F(\theta',x,t) \Delta t) \right) \rho (\theta',x,t) = 
\int_{0}^{2\pi} d\theta' \delta \left( a \theta - F(\theta,x,t) \Delta t - a \theta') \right) \rho (\theta',x,t) \, ,
\end{equation}
where we neglected terms ${\cal{O}}(\varepsilon^2)$ in the expansion, and defined $a \equiv 1 + \partial_\theta F(\theta,x,t) \Delta t$.
To perform the integration of the Dirac delta distribution, we change variables to $\phi = a \theta'$,
and approximate $a^{-1} \approx (1 - \partial_\theta F(\theta,x,t) \Delta t)$ to obtain
%
\begin{equation}
\int_{0}^{2\pi} d\theta' \delta \left( \theta - (\theta'+F(\theta',x,t) \Delta t) \right) \rho (\theta',x,t) =  
\rho(\theta,x,t) - \frac{\partial}{\partial\theta} \left( F(\theta,x,t) \rho(\theta,x,t) \right) \Delta t + {\cal{O}}(\Delta t^2) \, .
\end{equation}
Using this result in the gain term,
\begin{align}
\int_0^{2\pi} d\theta' P\left(\theta, x, t+\Delta t | \theta', x, t \right)\rho(\theta', x, t) = &
(1-\lambda \Delta t) \rho(\theta, x, t) - \frac{\partial}{\partial \theta} \left( F(\theta, x, t) \rho(\theta, x, t) \right) \Delta t  \nonumber \\
& + \lambda \Delta t\int_0^L dx' M(x'-x) \rho(\theta, x', t) + {\cal O}(\Delta t^2).
\end{align}
For the loss term of Eq.~(\ref{eq.master}), substitution of Eq.~(\ref{eq.transition_gain}) leads to
\begin{align}
\int_0^{2\pi} d\theta' P\left(\theta', x, t+\Delta t | \theta, x, t \right)\rho(\theta, x, t) = (1-\lambda \Delta t)\rho(\theta, x, t) +  \lambda \Delta t\rho(\theta, x, t).
\end{align}
Taking the limit $\Delta t \to 0$, we obtain the dynamic equation for the probability density
\begin{equation}
\frac{\partial \rho(\theta, x,t)}{\partial t} = - \frac{\partial}{\partial \theta} \left[ F(\theta,x, t) \rho(\theta, x, t) \right] +  \lambda \left[ \int_0^L dx' M(x'-x) \rho(\theta, x', t) - \rho(\theta, x, t) \right].
\label{eq.master_preexpan_ap}
\end{equation}

\section{Moment expansion of mobility kernel in the bulk}
For a short range mobility kernel, the mobility integral in Eq.~(\ref{eq.master_preexpan_ap}) involves contributions from short distances $|x'-x| \ll L$. 
Then, we can expand the probability density inside the integral around $x$,
\begin{equation}
\rho(\theta,x',t) = \rho(\theta,x,t) + \sum_{n=1}^{\infty} \frac{1}{n!} \frac{\partial^n}{\partial x^n} \rho(\theta,x,t) (x'-x)^n  \, .
\end{equation}
Using this expansion, the mobility integral becomes
\begin{equation}
\int_{0}^{L} dx' M(x'-x) \rho(\theta,x',t) = 
\rho(\theta,x,t) \int_{0}^{L} dx' M(x'-x) + 
\sum_{n=1}^{\infty} \frac{1}{n!} \frac{\partial^n}{\partial x^n} \rho(\theta,x,t) \int_{0}^{L} dx' (x'-x)^n M(x'-x)  \, .
\end{equation}
In the bulk, odd moments of the even kernel $M$ vanish and even moments are constant in space, $M_{2n}(x) \equiv M_{2n}$, see Appendix A.
Then replacing the mobility integral in the equation for the probability density Eq.~(\ref{eq.master_preexpan_ap}),
%
%
%
%
%
%
%
\begin{equation}
\frac{\partial \rho(\theta, x,t)}{\partial t} = - \frac{\partial}{\partial \theta} \left[ F(\theta, x, t) \rho(\theta, x, t) \right] +  \lambda \sum_{n=1}^{\infty} \frac{M_{2n}}{(2n)!} \frac{\partial^{2n} \rho(\theta, x, t)}{\partial x^{2n}}.
\label{eq.rho_cont_fullmoment}
\end{equation}
%
%
%

\section{Average phase}
In the main text, we define the average phase in a rotating reference frame $\omega = 0$ as
\begin{equation}
{\bar \theta(x, t)}  \equiv \int_{0}^{2\pi} d\theta \theta \rho(\theta, x, t).
\label{eq.def_ave_phase}
\end{equation}
This definition is meaningful if phase values are constrained in the interval $\theta \in [0, 2\pi)$.
%
%
%
We take the time derivative of Eq.~(\ref{eq.def_ave_phase}),
\begin{equation}
\frac{\partial {\bar \theta}(x, t)}{\partial t}  = \int_{0}^{2\pi} d\theta \theta \frac{\partial \rho(\theta, x, t)}{\partial t}.
\label{eq.der_ave_phase_ap}
\end{equation}
%
%
Substituting Eq.~(\ref{eq.rho_cont_fullmoment}) into Eq.~(\ref{eq.der_ave_phase_ap}) and integrating by parts with the vanishing surface terms
\begin{equation}
2 \pi \cdot F(2 \pi, x, t)\rho(2 \pi, x, t) = 0 \cdot F(0, x, t) \rho(0, x, t) = 0  \, ,
\end{equation}
we obtain
\begin{eqnarray}
\frac{\partial {\bar \theta}(x, t) }{\partial t}  =   \int_{0}^{2\pi} d\theta F(\theta, x, t)\rho(\theta, x, t)  + \lambda \sum_{n=1}^{\infty} \frac{M_{2n}}{(2n)!} \frac{\partial^{2n} {\bar \theta}(x, t)}{\partial x^{2n}} .
\label{eq.avephase3}
\end{eqnarray}
%
%
%
%
%
For a short ranged coupling kernel $r_c/L \ll 1$ and small variance probability densities, we can expand the sinusoidal function in $F(\theta,x,t)$, Eq.~(\ref{eq.F}), as $\sin (\theta' - \theta) = \theta' - \theta + {\cal O}(|\theta' - \theta|^3)$.
Then, we evaluate the integral in Eq.~(\ref{eq.avephase3}) and obtain
%
%
%
%
\begin{equation}
\frac{\partial {\bar \theta}(x, t) }{\partial t}  =  \kappa \int_0^L dx' C(x'-x) \left( {\bar \theta}(x', t) - {\bar \theta}(x, t) \right) + \lambda \sum_{n=1}^{\infty} \frac{M_{2n}}{(2n)!} \frac{\partial^{2n} {\bar \theta}(x, t)}{\partial x^{2n}}  + {\cal O}(|\theta' - \theta|^3)  \, ,
\label{eq.mean_full}
\end{equation}
where we used the marginal distribution definition $\int_{0}^{2\pi} d\theta \rho(\theta', x', t ; \theta, x, t) = \rho(\theta', x', t)$ 
and $\int_{0}^{2\pi} d\theta' \rho(\theta', x', t ; \theta, x, t) = \rho(\theta,x, t)$.
%
%
Next, we expand ${\bar \theta}(x', t)$ around $x$ and write Eq.~(\ref{eq.mean_full}) in terms of the coupling kernel moments.
Similarly to the mobility kernel $M(x)$ above, in the bulk, the odd moments of $C(x)$ vanish and the even moments are constants $C_{2n}(x) \equiv C_{2n}$ (Appendix A), so we obtain
\begin{equation}
\frac{\partial {\bar \theta}(x, t) }{\partial t}  =  \kappa \sum_{n=1}^{\infty} \frac{C_{2n}}{(2n)!} \frac{\partial^{2n} {\bar \theta}(x, t)}{\partial x^{2n}} +  \lambda \sum_{n=1}^{\infty} \frac{M_{2n}}{(2n)!} \frac{\partial^{2n} {\bar \theta}(x, t)}{\partial x^{2n}} + {\cal O}(|\theta' - \theta|^3) \, .
\end{equation}
%
%
%
Finally, neglecting higher order terms and moments we obtain the equation for the average phase
\begin{equation}
\frac{\partial {\bar \theta}(x, t) }{\partial t}  =  \frac{ \kappa C_2  + \lambda M_2 }{2}  \frac{\partial^2 {\bar \theta}(x, t)}{\partial x^2}.
\label{eq.pde_ave_ap}
\end{equation}

\section{Phase variance}
%
%
%
Next, we derive the time evolution of the phase variance
\begin{equation}
\sigma^2 (x,t) \equiv \int_{0}^{2\pi}d\theta \left( \theta - \bar{\theta}(x, t) \right)^2 \rho(\theta, x, t) \, .
\label{eq.var1}
\end{equation}
%
%
%
Expansion of the square term in Eq.~(\ref{eq.var1}) leads to
\begin{equation}
\sigma^2 (x,t) = \int_{0}^{2\pi}d\theta \theta^2 \rho(\theta, x, t) - \bar{\theta}(x,t)^2.
\label{eq.var2}
\end{equation}
The first term in Eq.~(\ref{eq.var2}) is the second moment of the phase $\langle \theta(x,t)^2 \rangle = \int_{0}^{2\pi}d\theta \theta^2 \rho(\theta, x, t)$.
Then, the time derivative of the variance is
\begin{equation}
\frac{\partial \sigma^2 (x,t)}{\partial t} = \int_{0}^{2\pi}d\theta \theta^2 \frac{\partial \rho(\theta, x, t)}{\partial t} - 2 \bar{\theta}(x,t)\frac{\partial \bar{\theta}(x,t)}{\partial t}.
\label{eq.var_dot1}
\end{equation}
In the first term, we substitute Eq.~(\ref{eq.rho_cont_fullmoment}) keeping up to second moment in the mobility kernel expansion.
Following the assumption we introduced for the average phase, we consider that phase fluctuations are relatively small.
We integrate by parts the term describing deterministic oscillator dynamics, assuming as above that the surface terms vanish $\int_0^{2\pi} d \theta \partial_\theta[\theta^2 F(\theta,x,t)\rho(\theta,x,t)] = 0$, and we exchange integration and derivation in the term describing mobility,
\begin{equation}
\int_{0}^{2\pi}d\theta \theta^2 \frac{\partial \rho(\theta, x, t)}{\partial t} = 2 \int_0^{2\pi} d\theta \theta F(\theta,x,t)\rho(\theta,x,t) + \frac{\lambda M_2}{2}\frac{\partial^2 \langle \theta(x,t)^2\rangle}{\partial x^2}.
\end{equation}
%
%
%
Next, we use the expression for $F(\theta, x,t)$ in Eq.~(\ref{eq.F}) with $\omega=0$. 
Expanding the sinusoidal function, we obtain
\begin{align}
\int_{0}^{2\pi}d\theta \theta^2 \frac{\partial \rho(\theta, x, t)}{\partial t} = 2\kappa \int_0^{2\pi} d\theta & \int_0^{2\pi}  d\theta' \int_0^L dx' C(x'-x) \,\, \theta \theta' \rho(\theta', x', t; \theta, x, t) \nonumber \\ &- 2 \kappa \langle \theta(x, t)^2 \rangle + \frac{\lambda M_2}{2}\frac{\partial^2 \langle \theta(x,t)^2\rangle}{\partial x^2} + {\cal O}(\theta \cdot |\theta'-\theta|^3).
\label{eq.var_first}
\end{align}
We define the phase covariance 
\begin{equation}
\mathrm{cov}(x,x',t) \equiv  \int_0^{2\pi} d\theta \int_0^{2\pi}  d\theta' (\theta - {\bar \theta}(x, t)) (\theta' - {\bar \theta}(x', t)) \rho(\theta',x',t; \theta,x,t).
\label{eq.cov}
\end{equation}
Using this covariance, we write
\begin{equation}
\int_0^{2\pi} d\theta \int_0^{2\pi}  d\theta' \theta \theta' \rho(\theta', x', t; \theta, x, t) = \mathrm{cov}(x, x', t) + {\bar \theta}(x',t){\bar \theta}(x,t).
\end{equation}
Then, after some calculation, Eq.~(\ref{eq.var_first}) becomes
\begin{align}
\int_{0}^{2\pi}d\theta \theta^2 \frac{\partial \rho(\theta, x, t)}{\partial t} = 2\kappa \int_0^L dx' &C(x'-x) \mathrm{cov}(x,x',t) -2\kappa \sigma^2(x,t)\nonumber\\ &+ \kappa C_2 {\bar \theta}(x,t) \frac{\partial^2 {\bar \theta}(x,t)}{\partial x^2} + \frac{\lambda M_2}{2} \frac{\partial^2 \langle \theta(x,t)^2\rangle}{\partial x^2}. 
\end{align}
In the second term of Eq.~(\ref{eq.var_dot1}), we substitute Eq.~(\ref{eq.pde_ave_ap}) and obtain
\begin{equation}
2 {\bar \theta}(x,t) \frac{\partial {\bar \theta}(x,t)}{\partial t} = \kappa C_2 {\bar \theta}(x,t) \frac{\partial^2 {\bar \theta}(x,t)}{\partial x^2} + \lambda M_2 {\bar \theta}(x,t) \frac{\partial^2 {\bar \theta}(x,t)}{\partial x^2}.
\end{equation}
Putting together these results, the time derivative of variance can be written as
\begin{align}
\frac{\partial \sigma^2 (x,t)}{\partial t} &= \int_{0}^{2\pi}d\theta \theta^2 \frac{\partial \rho(\theta, x, t)}{\partial t} - 2 \bar{\theta}(x,t)\frac{\partial \bar{\theta}(x,t)}{\partial t} \nonumber \\
& = 2\kappa \int_0^L dx' C(x'-x) \mathrm{cov}(x,x',t) -2\kappa \sigma^2(x,t) + \frac{\lambda M_2}{2} \frac{\partial^2 \langle \theta(x,t)^2\rangle}{\partial x^2} - \lambda M_2 {\bar \theta}(x,t) \frac{\partial^2 {\bar \theta}(x,t)}{\partial x^2}.
\end{align}
%
%
Finally, from the chain rule for ${\partial^2 \bar{\theta}^2}/{\partial x^2}$, we can use the relation
\begin{equation}
{\bar \theta}(x,t) \frac{\partial^2 {\bar \theta}(x,t)}{\partial x^2} = \frac{1}{2} \frac{\partial^2 {\bar \theta}(x,t)^2}{\partial x^2} - \left( \frac{\partial {\bar \theta}(x,t)}{\partial x} \right)^2 \, ,
\end{equation}
and obtain
\begin{align}
\frac{\partial \sigma^2 (x,t)}{\partial t} = 2\kappa \int_0^L dx' C(x'-x) \mathrm{cov}(x,x',t) -2\kappa \sigma^2(x,t) + \frac{\lambda M_2}{2} \frac{\partial^2 \sigma^2 (x,t)}{\partial x^2} + \lambda M_2 \left( \frac{\partial {\bar \theta}(x,t)}{\partial x} \right)^2.
\end{align}

\section{Statistical description in higher dimensions}
In this section we consider the statistical description of mobile oscillators in two-dimensional lattice spaces.
%
%
We briefly outline the derivation since it parallels that for the one-dimensional model described above.
The derivation may be extended to higher dimensions as well.
We begin with the phase equation for mobile oscillators in a two-dimensional lattice

\begin{equation} \label{eq.phase_disc_2D}
\dot\theta_{ij}(t) = \omega + \kappa \sum_{i', j'=0}^{N} {C_{ij,i'j'} \sin\left(\theta_{i'j'}(t) - \theta_{ij}(t)\right)} \, ,
\end{equation}
where the indices $i$ and $j$ label lattice sites along the two spatial coordinates $x$ and $y$, with corresponding positions $x_i=i \Delta x$ and $y_j = j \Delta y$.
The coupling kernel is 
$C_{ij,i'j'} = C({\bf x}, {\bf x'})\Delta x \Delta y$, 
where we introduced the vector notation ${\bf x}=(x,y)$.
We assume a symmetric kernel, $C({\bf x}, {\bf x'}) = C(|{\bf x'}-{\bf x}|)$.
We take the limit $\Delta x \to 0$, $\Delta y \to 0$ and write the sum as an integral, 
\begin{equation}
\frac{\partial \theta({\bf x}, t)}{\partial t} = \omega + \kappa \int_0^L dx' \int_0^L dy' C(|{\bf x}'-{\bf x}|) \sin \left( \theta({\bf x}',t) - \theta({\bf x},t) \right) \, .
\label{eq.phase_cont_2D}
\end{equation}
We next consider a large ensemble of independent systems, and following the previous derivations for the one-dimensional case we obtain the dynamics for the probability density,

\begin{equation}
\frac{\partial \rho(\theta, {\bf x},t)}{\partial t} = - \frac{\partial}{\partial \theta} \left[ F(\theta,{\bf x}, t) \rho(\theta, {\bf x}, t) \right] +  \lambda \left[ \int_0^L dx' \int_0^L dy' M(|{\bf x}'-{\bf x}|) \rho(\theta, {\bf x}', t) - \rho(\theta, {\bf x}, t) \right] \,,
\label{eq.master_preexpan_2D}
\end{equation}
where
\begin{equation}
F\left(\theta, {\bf x},t\right) = \omega + \kappa \int_0^{2\pi} d\theta' \int_0^L dx' \int_0^L dy' C(|{\bf x}'-{\bf x}|) \sin \left(\theta' - \theta \right) \rho(\theta', {\bf x}', t | \theta, {\bf x}, t) \, ,
\label{eq.F_2D}
\end{equation}
and we assumed a symmetric mobility kernel $M({\bf x}, {\bf x'}) = M(|{\bf x'}-{\bf x}|)$.

Next, we expand the probability density inside the mobility integral in Eq.~(\ref{eq.master_preexpan_2D}) around ${\bf x}$,
\begin{equation}
\rho(\theta,{\bf x}',t) = \sum_{n=0}^{\infty} \sum_{m=0}^{\infty} \frac{1}{n!m!} \frac{\partial^{n+m}}{\partial x^n \partial y^m} \rho(\theta,{\bf x},t) (x'-x)^n (y'-y)^m  \, .
\end{equation}
We use this expansion to evaluate the mobility integral,
\begin{equation}
\int_0^L dx' \int_0^L dy' M(|{\bf x}'-{\bf x}|) \rho(\theta, {\bf x}', t) = 
\sum_{n=0}^{\infty} \sum_{m=0}^{\infty} \frac{1}{n!m!} \frac{\partial^{n+m}}{\partial x^n \partial y^m} \rho(\theta,{\bf x},t) \int_{0}^{L} dx' \int_{0}^{L} dy' M(|{\bf x}'-{\bf x}|) (x'-x)^n (y'-y)^m  \, .
\end{equation}
The symmetry of the mobility kernel makes it an even function of both coordinates, so the kernel moments vanish for both odd $n$ and odd $m$.
Introducing moments
\begin{equation}
M_{nm} = \int_{0}^{L} dx' \int_{0}^{L} dy' M(|{\bf x}'-{\bf x}|) (x'-x)^n (y'-y)^m  \, ,
\end{equation}
\begin{equation}
\int_0^L dx' \int_0^L dy' M(|{\bf x}'-{\bf x}|) \rho(\theta, {\bf x}', t) = 
\sum_{n=0}^{\infty} \sum_{m=0}^{\infty} \frac{M_{2n \, 2m}}{2n!\,2m!} \frac{\partial^{2n+2m}}{\partial x^{2n} \partial y^{2m}} \rho(\theta,{\bf x},t) \, .
\end{equation}
We keep up to second order terms to obtain an approximate expression for the mobility integral,
\begin{equation}
\int_0^L dx' \int_0^L dy' M(|{\bf x}'-{\bf x}|) \rho(\theta, {\bf x}', t) = 
\rho(\theta,{\bf x},t) + \frac{M_2}{2} \nabla^2 \rho(\theta,{\bf x},t) \, ,
\end{equation}
where $M_{02} = M_{20} \equiv M_2$ from the symmetry of the mobility kernel.
Using this result back into Eq.~(\ref{eq.master_preexpan_2D}), 
\begin{equation}
\frac{\partial \rho(\theta, {\bf x},t)}{\partial t} = - \frac{\partial}{\partial \theta} \left[ F(\theta,{\bf x}, t) \rho(\theta, {\bf x}, t) \right] +  \frac{\lambda M_2}{2} \nabla^2 \rho(\theta,{\bf x},t) \,.
\label{eq.master_expan_2D}
\end{equation}

We define the average phase at position ${\bf x}$ and time $t$ in a rotating reference frame $\omega = 0$ as
\begin{equation}
{\bar \theta({\bf x}, t)}  \equiv \int_{0}^{2\pi} d\theta \theta \rho(\theta, {\bf x}, t).
\label{eq.def_ave_phase_2D}
\end{equation}
Taking the time derivative, we obtain
\begin{eqnarray}
\frac{\partial {\bar \theta}({\bf x}, t) }{\partial t}  =   \int_{0}^{2\pi} d\theta F(\theta, {\bf x}, t)\rho(\theta, {\bf x}, t)  + \frac{\lambda M_2}{2} \nabla^2 {\bar \theta}({\bf x},t) \, .
\label{eq.avephase3_2D}
\end{eqnarray}
where surface terms after integration by parts vanish assuming that phase values are constrained in the interval $[0,2\pi)$.
For a short ranged coupling kernel $r_c/L \ll 1$ and small variance probability densities, we can expand the sinusoidal function in $F(\theta,{\bf x},t)$, as $\sin (\theta' - \theta) = \theta' - \theta + {\cal O}(|\theta' - \theta|^3)$.
Then, we evaluate the integral in Eq.~(\ref{eq.avephase3_2D}) and obtain
\begin{equation}
\frac{\partial {\bar \theta}({\bf x}, t) }{\partial t}  =  \kappa \int_0^L dx' \int_0^L dy' C(|{\bf x}'-{\bf x}|) \left( {\bar \theta}({\bf x}', t) - {\bar \theta}({\bf x}, t) \right) + \frac{\lambda M_2}{2} \nabla^2 {\bar \theta}({\bf x},t) + {\cal O}(|\theta' - \theta|^3)  \, ,
\label{eq.mean_full_2D}
\end{equation}
Next, we expand ${\bar \theta}({\bf x}', t)$ around ${\bf x}$.
Due to the symmetry of the coupling kernel, odd moments vanish similar to the mobility kernel expansion.
Keeping up to second order, 
\begin{equation}
\frac{\partial {\bar \theta}({\bf x}, t) }{\partial t} = \frac{ \kappa C_2  + \lambda M_2 }{2}  \nabla^2 {\bar \theta}({\bf x}, t) \, .
\label{eq.pde_ave_2D}
\end{equation}
where $C_2 \equiv C_{02} = C_{20}$.

Next, we define phase variance
\begin{equation}
\sigma^2 ({\bf x},t) \equiv \int_{0}^{2\pi}d\theta \left( \theta - \bar{\theta}({\bf x}, t) \right)^2 \rho(\theta, {\bf x}, t) \, .
\label{eq.var_2D}
\end{equation}
The time derivative of the variance is
\begin{equation}
\frac{\partial \sigma^2 ({\bf x},t)}{\partial t} = \int_{0}^{2\pi}d\theta \theta^2 \frac{\partial \rho(\theta, {\bf x}, t)}{\partial t} - 2 \bar{\theta}({\bf x},t)\frac{\partial \bar{\theta}({\bf x},t)}{\partial t}.
\label{eq.var_dot1_2D}
\end{equation}
Introducing the phase covariance in two-dimensional space 
\begin{equation}
\mathrm{cov}({\bf x},{\bf x}',t) \equiv  \int_0^{2\pi} d\theta \int_0^{2\pi}  d\theta' (\theta - {\bar \theta}({\bf x}, t)) (\theta' - {\bar \theta}({\bf x}', t)) \rho(\theta',{\bf x}',t; \theta,{\bf x},t) \, ,
\label{eq.cov_2D}
\end{equation}
and following the steps in one-dimensional case calculation, we obtain the variance equation
\begin{align}
\frac{\partial \sigma^2 ({\bf x},t)}{\partial t} = 
2\kappa \int_0^L dx' \int_0^L dy'
C(|{\bf x}'-{\bf x}|) \mathrm{cov}({\bf x},{\bf x}',t) -2\kappa \sigma^2({\bf x},t) + \frac{\lambda M_2}{2} \nabla^2 \sigma^2 ({\bf x},t) + \lambda M_2 \left( \nabla {\bar \theta}({\bf x},t) \right)^2 \,.
\label{eq.var_dot2_2D}
\end{align}
Assuming independence of events in the joint probability density, we can neglect the covariance convolution similar to the one-dimensional case, 
\begin{equation}
\frac{\partial \sigma^2 ({\bf x},t)}{\partial t} \approx -2\kappa \sigma^2({\bf x},t) + \frac{\lambda M_2}{2} \nabla^2 \sigma^2 ({\bf x},t) + \lambda M_2 \left( \nabla {\bar \theta}({\bf x},t) \right)^2 \,.
\label{eq.var_dot_approx_2D}
\end{equation}
In Fig.~\ref{fig:2d_example}, we plot simulation results of Eq.~(\ref{eq.phase_disc_2D}) together with numerical solutions of Eqs.~(\ref{eq.pde_ave_2D}) and (\ref{eq.var_dot_approx_2D}) by employing nearest neighbor coupling and mobility kernels.


\section{Mean-field transition}
Here we derive a transition point where spatial modes cannot grow from random initial phases due to fast mobility $\lambda / \kappa \gg 1$. 
%
Such a transition point marks the onset of mean-field behavior of the mobile oscillators.
%
%
%
With this fast mobility, phase correlations between neighboring positions are expected to be small, and the joint probability density implicit in Eq.~(\ref{eq.master_preexpan}) may be approximated as
\begin{equation}
\rho(\theta', x', t; \theta, x, t) \approx \rho(\theta', x', t) \rho(\theta, x, t).
\end{equation}
We express the $2 \pi$-periodic function $\rho(\theta, x, t)$ using Fourier series for $\theta$,
\begin{equation}
\rho(\theta, x, t) = \frac{1}{2 \pi} + \sum_{l \neq 0} \eta_l (x, t) e^{i l \theta},
\label{eq.rho_fourier_ap}
\end{equation}
which satisfies the normalization condition $\int_{0}^{2 \pi} d\theta \rho(\theta, x, t) = 1$.
Substituting Eq.~(\ref{eq.rho_fourier_ap}) into Eq.~(\ref{eq.rho_cont_fullmoment}), keeping up to second order in mobility moment expansion, and collecting terms including $e^{i l \theta}$, we obtain 
\begin{align}
\frac{\partial \eta_{1}(x,t)}{\partial t} =  \frac{{\bar \lambda}}{2} \frac{\partial^2 \eta_{1}(x,t)}{\partial x^2} - i\omega \eta_{1}(x,t) - \kappa \int_{0}^{L} dx' C(x'-x) \left( -\frac{1}{2}\eta_{1}(x',t) + \pi \eta_{-1}(x',t) \eta_{2}(x,t)  \right) \, ,
\label{eq.etaplus_nonline}
\end{align}
\begin{align}
\frac{\partial \eta_{-1}(x,t)}{\partial t} =  \frac{{\bar \lambda}}{2} \frac{\partial^2 \eta_{-1}(x,t)}{\partial x^2} + i\omega \eta_{-1}(x,t) - \kappa \int_{0}^{L} dx' C(x'-x) \left( -\frac{1}{2}\eta_{-1}(x',t) + \pi \eta_{1}(x',t) \eta_{-2}(x,t)  \right) ,
\label{eq.etaminus_nonline}
\end{align}
for $l=\pm 1$, and
\begin{align}
\frac{\partial \eta_{l}(x,t)}{\partial t} =  \frac{{\bar \lambda}}{2} \frac{\partial^2 \eta_{l}(x,t)}{\partial x^2} - i \omega l \eta_{l}(x,t) - \kappa \pi l \int_{0}^{L} dx' C(x'-x) \left( \eta_{-1}(x',t) \eta_{l+1}(x,t) - \eta_{1}(x',t) \eta_{l-1}(x,t) \right),
\label{eq.etal_nonline}
\end{align}
for $l \neq \pm 1$.
%
%
%
%
%
%
Thus, for $|l|>1$, coupling mediated $\eta_l$ growth depends on second order terms including $\eta_{\pm 1}$.
Hence, we expect that $\eta_{\pm 1}$ grow first and then $\eta_l$ follows.
If we neglect the nonlinear term in Eq.~(\ref{eq.etaminus_nonline}) assuming $\eta_l(x, t) \ll 1$ at early times, we obtain
\begin{align}
\frac{\partial \eta_{-1}(x,t)}{\partial t} \approx  \frac{{\bar \lambda}}{2} &\frac{\partial^2 \eta_{-1}(x,t)}{\partial x^2} + i\omega \eta_{-1}(x,t) + \frac{\kappa}{2} \int_0^L dx' C(x'-x) \eta_{-1}(x',t).
\label{eq.etaminus_line}
\end{align}
A similar calculation applies to $\eta_{1}(x,t)$.

To examine the growth of spatially uniform component of $\eta_{-1}$ in Eq.~(\ref{eq.etaminus_line}), we assume $\eta_{-1}(x, t) = \epsilon_h e^{\mu_h t}$ where $\mu_h$ is the growth rate of the spatially uniform component and $\epsilon_h$ is a constant determined by the initial condition.
Substitution of this expression into Eq.~(\ref{eq.etaminus_line}) leads to $\mu_h = i \omega + \kappa /2$.
Hence, the real part of $\mu_h$ is $\kappa /2 > 0$, indicating that the spatially uniform component of $\eta_{-1}$ grows over time due to coupling.

We then consider the following spatio-temporal dependence of $\eta_{-1}$ based on the open boundary condition imposed on the mobile oscillators,
\begin{equation}
\eta_{-1}(x, t) \approx \epsilon e^{\mu t} \cos \left( \frac{\pi}{L} x \right).
\label{eq.eta_cos}
\end{equation}
Substitution of Eq.~(\ref{eq.eta_cos}) into Eq.~(\ref{eq.etaminus_line}), and truncation of higher moments lead to the growth rate of the spatial mode
\begin{equation}
\mu \approx -\frac{{\bar \lambda}}{2} \left( \frac{\pi }{L} \right)^2 + \frac{\kappa}{2} -\frac{{\bar \kappa}}{4} \left( \frac{\pi }{L} \right)^2  + i\omega,
\label{eq.mu}
\end{equation}
where ${\bar \lambda} = M_2 \lambda$ and ${\bar \kappa} = C_2 \kappa$.
We argue that the mean-filed behavior emerges when this longest spatial mode does not grow due to fast mobility.
This argument provides the onset of mean-filed behavior as ${\rm Re}[\mu] = 0$, which occurs at
\begin{equation}
\frac{{\bar \lambda}}{\kappa} \approx \left( \frac{L}{\pi} \right)^2 -\frac{C_2}{2} \approx \left( \frac{L}{\pi} \right)^2.
\label{eq.mfonset_ap}
\end{equation}
Thus, the onset of mean-field behavior is proportional to $L^2$.
%

\section{Statistical description of phase waves in vertebrate segment formation}
Here we derive a statistical description in the presence of oscillator advection, and frequency and mobility gradients.
%
%
We begin with Eq.~(15) of the main text for mobile oscillators in a discrete lattice,
\begin{align}
\frac{d \theta_i(t)}{dt} = \omega_i + v_a (\theta_{i+1}(t) - \theta_{i}(t)) + \frac{\kappa}{n_i} \sum_{j = i \pm 1} \sin \left(\theta_j(t) - \theta_i(t) \right) \, ,
\label{eq.fgad_orig_ap}
\end{align}
where $\theta_i(t)$ is the phase at site $i$ and time $t$, and $\omega_i$ is the frequency gradient.
%
%
%
%
%
%
The second term in Eq.~(\ref{eq.fgad_orig_ap}) represents the advection of oscillators due to the axis elongation of embryos and $v_a$ is the advection speed.
For simplicity, we assume that $v_a$ is constant over time and space.
The third term represents coupling between oscillators at neighboring lattice sites.

Embryonic patterns often display more than one gene expression wave along the tissue.
These multiple waves may be represented by phase values spanning a range beyond $2\pi$.
Therefore, here we extend the phase space to the range $[-2n\pi,2n\pi]$, and consider a large $n$ value so that the support of probability density $\rho(\theta, x, t)$ remains in this range at all times.
%
%
Following the homogeneous space derivation, we obtain the evolution equation for the probability density up to the second order in $\Delta x$,
%
%
\begin{eqnarray}
\frac{\partial \rho(\theta, x, t)}{\partial t} = -\frac{\partial}{\partial \theta} \left[ G(\theta, x, t) \rho(\theta, x, t) \right] + \frac{\lambda(x) \Delta x^2}{2} \frac{\partial^2 \rho(\theta, x, t)}{\partial x^2},
\label{eq.pb1}
\end{eqnarray}
where
\begin{align}
G(\theta,x,t) & = \omega(x) + v_a \int_{-2n\pi}^{2n\pi} d\theta' \left(\theta' - \theta \right) \rho(\theta', x+\Delta x, t | \theta, x, t) \nonumber \\
& + \frac{\kappa}{2} \int_{-2n\pi}^{2n\pi} d\theta' \sin(\theta'-\theta) \rho(\theta', x+\Delta x, t | \theta, x, t) + \frac{\kappa}{2} \int_{-2n\pi}^{2n\pi} d\theta' \sin(\theta'-\theta) \rho(\theta', x-\Delta x, t | \theta, x, t).
\label{eq.F0_a}
\end{align}
The deterministic drift $G(\theta, x, t)$ now includes the advection term.
%
%
The extension of the integration range does not affect calculations introduced in the previous sections.

%
%
We then compute the time evolution of average phase integrating $G$ term by parts, with surface terms vanishing at the boundaries $\pm 2n\pi$, and obtain
\begin{equation}
\frac{\partial  {\bar \theta(x, t)} }{\partial t}  = \omega(x) + v_a \frac{\partial {\bar \theta}(x,t)}{\partial x} \Delta x 
+  \frac{\kappa + \lambda(x)}{2} \frac{\partial^2  {\bar \theta}(x, t)}{\partial x^2}\Delta x^2 + {\cal O}(v_a\Delta x^2) \, .
\end{equation}
Introducing the scaling
\begin{eqnarray} \label{eq.va_scaling}
v_a &=& {\bar v}_a/ \Delta x \,, \nonumber \\
\kappa &=& {\bar \kappa}/ \Delta x^2 \,, \\
\lambda(x) &=& {\bar \lambda}(x)/ \Delta x^2 \, , \nonumber 
\end{eqnarray}
and taking the limit $\Delta x \to 0$, 
%
%
\begin{equation}
\frac{\partial  {\bar \theta(x, t)} }{\partial t}  = \omega(x) + {\bar v}_a \frac{\partial {\bar \theta}(x,t)}{\partial x} 
+   \frac{{\bar \kappa} + {\bar \lambda}(x)}{2} \frac{\partial^2  {\bar \theta}(x, t)}{\partial x^2}.
\label{eq.qbar}
\end{equation}

Similarly, we obtain the time evolution of the variance, after neglecting covariance terms, 
%
%
%
\begin{align}
\frac{\partial \sigma^2 (x,t)}{\partial t} \approx - 2 \left(\frac{{\bar \kappa}}{\Delta x^2} + \frac{{\bar v}_a}{\Delta x} \right) \sigma^2(x,t) + \frac{{\bar \lambda}(x)}{2} \frac{\partial^2 \sigma^2 (x,t)}{\partial x^2} +  {\bar \lambda}(x) \left( \frac{\partial \bar{\theta}(x,t)}{\partial x} \right)^2,
\label{eq.var_fgncv_ap}
\end{align}
where we used the scaled parameters defined in Eq.~(\ref{eq.va_scaling}).
The first term in Eq.~(\ref{eq.var_fgncv_ap}) indicates that variance is reduced by both coupling and advection.
Note that this term includes the lattice space $\Delta x$.
The second and third terms represent diffusion and production of variance, respectively, accompanied by the movement of oscillators.

\twocolumngrid

\bibliography{statistical_description}

\begin{thebibliography}{64}%
\makeatletter
\providecommand \@ifxundefined [1]{%
 \@ifx{#1\undefined}
}%
\providecommand \@ifnum [1]{%
 \ifnum #1\expandafter \@firstoftwo
 \else \expandafter \@secondoftwo
 \fi
}%
\providecommand \@ifx [1]{%
 \ifx #1\expandafter \@firstoftwo
 \else \expandafter \@secondoftwo
 \fi
}%
\providecommand \natexlab [1]{#1}%
\providecommand \enquote  [1]{``#1''}%
\providecommand \bibnamefont  [1]{#1}%
\providecommand \bibfnamefont [1]{#1}%
\providecommand \citenamefont [1]{#1}%
\providecommand \href@noop [0]{\@secondoftwo}%
\providecommand \href [0]{\begingroup \@sanitize@url \@href}%
\providecommand \@href[1]{\@@startlink{#1}\@@href}%
\providecommand \@@href[1]{\endgroup#1\@@endlink}%
\providecommand \@sanitize@url [0]{\catcode `\\12\catcode `\$12\catcode
  `\&12\catcode `\#12\catcode `\^12\catcode `\_12\catcode `\%12\relax}%
\providecommand \@@startlink[1]{}%
\providecommand \@@endlink[0]{}%
\providecommand \url  [0]{\begingroup\@sanitize@url \@url }%
\providecommand \@url [1]{\endgroup\@href {#1}{\urlprefix }}%
\providecommand \urlprefix  [0]{URL }%
\providecommand \Eprint [0]{\href }%
\providecommand \doibase [0]{http://dx.doi.org/}%
\providecommand \selectlanguage [0]{\@gobble}%
\providecommand \bibinfo  [0]{\@secondoftwo}%
\providecommand \bibfield  [0]{\@secondoftwo}%
\providecommand \translation [1]{[#1]}%
\providecommand \BibitemOpen [0]{}%
\providecommand \bibitemStop [0]{}%
\providecommand \bibitemNoStop [0]{.\EOS\space}%
\providecommand \EOS [0]{\spacefactor3000\relax}%
\providecommand \BibitemShut  [1]{\csname bibitem#1\endcsname}%
\let\auto@bib@innerbib\@empty
\bibitem [{\citenamefont {Strogatz}(2004)}]{sync}%
  \BibitemOpen
  \bibfield  {author} {\bibinfo {author} {\bibfnamefont {S.~H.}\ \bibnamefont
  {Strogatz}},\ }\href@noop {} {\emph {\bibinfo {title} {Sync: How Order
  Emerges From Chaos In the Universe, Nature, and Daily Life}}},\ \bibinfo
  {edition} {1st}\ ed.\ (\bibinfo  {publisher} {Hyperion},\ \bibinfo {year}
  {2004})\BibitemShut {NoStop}%
\bibitem [{\citenamefont {Pikovsky}\ \emph {et~al.}(2001)\citenamefont
  {Pikovsky}, \citenamefont {Rosenblum},\ and\ \citenamefont
  {Kurths}}]{pikovsky}%
  \BibitemOpen
  \bibfield  {author} {\bibinfo {author} {\bibfnamefont {A.~S.}\ \bibnamefont
  {Pikovsky}}, \bibinfo {author} {\bibfnamefont {M.~G.}\ \bibnamefont
  {Rosenblum}}, \ and\ \bibinfo {author} {\bibfnamefont {J.}~\bibnamefont
  {Kurths}},\ }\href@noop {} {\emph {\bibinfo {title} {Synchronization: a
  Universal Concept in Nonlinear Sciences}}}\ (\bibinfo  {publisher} {Cambridge
  University Press},\ \bibinfo {address} {Cambridge},\ \bibinfo {year}
  {2001})\BibitemShut {NoStop}%
\bibitem [{\citenamefont {Taylor}\ \emph {et~al.}(2009)\citenamefont {Taylor},
  \citenamefont {Tinsley}, \citenamefont {Wang}, \citenamefont {Huang},\ and\
  \citenamefont {Showalter}}]{taylor09}%
  \BibitemOpen
  \bibfield  {author} {\bibinfo {author} {\bibfnamefont {A.~F.}\ \bibnamefont
  {Taylor}}, \bibinfo {author} {\bibfnamefont {M.~R.}\ \bibnamefont {Tinsley}},
  \bibinfo {author} {\bibfnamefont {F.}~\bibnamefont {Wang}}, \bibinfo {author}
  {\bibfnamefont {Z.}~\bibnamefont {Huang}}, \ and\ \bibinfo {author}
  {\bibfnamefont {K.}~\bibnamefont {Showalter}},\ }\href@noop {} {\bibfield
  {journal} {\bibinfo  {journal} {Science}\ }\textbf {\bibinfo {volume}
  {323}},\ \bibinfo {pages} {614} (\bibinfo {year} {2009})}\BibitemShut
  {NoStop}%
\bibitem [{\citenamefont {Uriu}\ \emph {et~al.}(2017)\citenamefont {Uriu},
  \citenamefont {Bhavna}, \citenamefont {Oates},\ and\ \citenamefont
  {Morelli}}]{uriu17b}%
  \BibitemOpen
  \bibfield  {author} {\bibinfo {author} {\bibfnamefont {K.}~\bibnamefont
  {Uriu}}, \bibinfo {author} {\bibfnamefont {R.}~\bibnamefont {Bhavna}},
  \bibinfo {author} {\bibfnamefont {A.~C.}\ \bibnamefont {Oates}}, \ and\
  \bibinfo {author} {\bibfnamefont {L.~G.}\ \bibnamefont {Morelli}},\
  }\href@noop {} {\bibfield  {journal} {\bibinfo  {journal} {Biology Open}\ ,\
  \bibinfo {pages} {bio}} (\bibinfo {year} {2017})}\BibitemShut {NoStop}%
\bibitem [{\citenamefont {Buscarino}\ \emph {et~al.}(2006)\citenamefont
  {Buscarino}, \citenamefont {Fortuna}, \citenamefont {Frasca},\ and\
  \citenamefont {Rizzo}}]{buscarino06}%
  \BibitemOpen
  \bibfield  {author} {\bibinfo {author} {\bibfnamefont {A.}~\bibnamefont
  {Buscarino}}, \bibinfo {author} {\bibfnamefont {L.}~\bibnamefont {Fortuna}},
  \bibinfo {author} {\bibfnamefont {M.}~\bibnamefont {Frasca}}, \ and\ \bibinfo
  {author} {\bibfnamefont {A.}~\bibnamefont {Rizzo}},\ }\href@noop {}
  {\bibfield  {journal} {\bibinfo  {journal} {Chaos: An Interdisciplinary
  Journal of Nonlinear Science}\ }\textbf {\bibinfo {volume} {16}},\ \bibinfo
  {pages} {015116} (\bibinfo {year} {2006})}\BibitemShut {NoStop}%
\bibitem [{\citenamefont {Uriu}\ \emph {et~al.}(2014)\citenamefont {Uriu},
  \citenamefont {Morelli},\ and\ \citenamefont {Oates}}]{uriu14b}%
  \BibitemOpen
  \bibfield  {author} {\bibinfo {author} {\bibfnamefont {K.}~\bibnamefont
  {Uriu}}, \bibinfo {author} {\bibfnamefont {L.~G.}\ \bibnamefont {Morelli}}, \
  and\ \bibinfo {author} {\bibfnamefont {A.~C.}\ \bibnamefont {Oates}},\
  }\href@noop {} {\bibfield  {journal} {\bibinfo  {journal} {Semin. Cell Dev.
  Biol.}\ }\textbf {\bibinfo {volume} {35}},\ \bibinfo {pages} {66} (\bibinfo
  {year} {2014})}\BibitemShut {NoStop}%
\bibitem [{\citenamefont {Lewis}(2003)}]{lewis03}%
  \BibitemOpen
  \bibfield  {author} {\bibinfo {author} {\bibfnamefont {J.}~\bibnamefont
  {Lewis}},\ }\href@noop {} {\bibfield  {journal} {\bibinfo  {journal} {Curr.
  Biol.}\ }\textbf {\bibinfo {volume} {13}},\ \bibinfo {pages} {1398} (\bibinfo
  {year} {2003})}\BibitemShut {NoStop}%
\bibitem [{\citenamefont {Schr{\"{o}}ter}\ \emph {et~al.}(2012)\citenamefont
  {Schr{\"{o}}ter}, \citenamefont {Ares}, \citenamefont {Morelli},
  \citenamefont {Isakova}, \citenamefont {Hens}, \citenamefont {Soroldoni},
  \citenamefont {Gajewski}, \citenamefont {J{\"{u}}licher}, \citenamefont
  {Maerkl}, \citenamefont {Deplancke},\ and\ \citenamefont
  {Oates}}]{schroter12}%
  \BibitemOpen
  \bibfield  {author} {\bibinfo {author} {\bibfnamefont {C.}~\bibnamefont
  {Schr{\"{o}}ter}}, \bibinfo {author} {\bibfnamefont {S.}~\bibnamefont
  {Ares}}, \bibinfo {author} {\bibfnamefont {L.~G.}\ \bibnamefont {Morelli}},
  \bibinfo {author} {\bibfnamefont {A.}~\bibnamefont {Isakova}}, \bibinfo
  {author} {\bibfnamefont {K.}~\bibnamefont {Hens}}, \bibinfo {author}
  {\bibfnamefont {D.}~\bibnamefont {Soroldoni}}, \bibinfo {author}
  {\bibfnamefont {M.}~\bibnamefont {Gajewski}}, \bibinfo {author}
  {\bibfnamefont {F.}~\bibnamefont {J{\"{u}}licher}}, \bibinfo {author}
  {\bibfnamefont {S.~J.}\ \bibnamefont {Maerkl}}, \bibinfo {author}
  {\bibfnamefont {B.}~\bibnamefont {Deplancke}}, \ and\ \bibinfo {author}
  {\bibfnamefont {A.~C.}\ \bibnamefont {Oates}},\ }\href@noop {} {\bibfield
  {journal} {\bibinfo  {journal} {PLoS Biology}\ }\textbf {\bibinfo {volume}
  {10}},\ \bibinfo {pages} {e1001364} (\bibinfo {year} {2012})}\BibitemShut
  {NoStop}%
\bibitem [{\citenamefont {Jiang}\ \emph {et~al.}(2000)\citenamefont {Jiang},
  \citenamefont {Aerne}, \citenamefont {Smithers}, \citenamefont {Haddon},
  \citenamefont {Ish-Horowicz},\ and\ \citenamefont {Lewis}}]{jiang00}%
  \BibitemOpen
  \bibfield  {author} {\bibinfo {author} {\bibfnamefont {Y.-J.}\ \bibnamefont
  {Jiang}}, \bibinfo {author} {\bibfnamefont {B.~L.}\ \bibnamefont {Aerne}},
  \bibinfo {author} {\bibfnamefont {L.}~\bibnamefont {Smithers}}, \bibinfo
  {author} {\bibfnamefont {C.}~\bibnamefont {Haddon}}, \bibinfo {author}
  {\bibfnamefont {D.}~\bibnamefont {Ish-Horowicz}}, \ and\ \bibinfo {author}
  {\bibfnamefont {J.}~\bibnamefont {Lewis}},\ }\href@noop {} {\bibfield
  {journal} {\bibinfo  {journal} {Nature}\ }\textbf {\bibinfo {volume} {408}},\
  \bibinfo {pages} {475} (\bibinfo {year} {2000})}\BibitemShut {NoStop}%
\bibitem [{\citenamefont {Horikawa}\ \emph {et~al.}(2006)\citenamefont
  {Horikawa}, \citenamefont {Ishimatsu}, \citenamefont {Yoshimoto},
  \citenamefont {Kondo},\ and\ \citenamefont {Takeda}}]{horikawa06}%
  \BibitemOpen
  \bibfield  {author} {\bibinfo {author} {\bibfnamefont {K.}~\bibnamefont
  {Horikawa}}, \bibinfo {author} {\bibfnamefont {K.}~\bibnamefont {Ishimatsu}},
  \bibinfo {author} {\bibfnamefont {E.}~\bibnamefont {Yoshimoto}}, \bibinfo
  {author} {\bibfnamefont {S.}~\bibnamefont {Kondo}}, \ and\ \bibinfo {author}
  {\bibfnamefont {H.}~\bibnamefont {Takeda}},\ }\href@noop {} {\bibfield
  {journal} {\bibinfo  {journal} {Nature}\ }\textbf {\bibinfo {volume} {441}},\
  \bibinfo {pages} {719} (\bibinfo {year} {2006})}\BibitemShut {NoStop}%
\bibitem [{\citenamefont {Riedel-Kruse}\ \emph {et~al.}(2007)\citenamefont
  {Riedel-Kruse}, \citenamefont {M{\"u}ller},\ and\ \citenamefont
  {Oates}}]{riedel07}%
  \BibitemOpen
  \bibfield  {author} {\bibinfo {author} {\bibfnamefont {I.~H.}\ \bibnamefont
  {Riedel-Kruse}}, \bibinfo {author} {\bibfnamefont {C.}~\bibnamefont
  {M{\"u}ller}}, \ and\ \bibinfo {author} {\bibfnamefont {A.~C.}\ \bibnamefont
  {Oates}},\ }\href@noop {} {\bibfield  {journal} {\bibinfo  {journal}
  {Science}\ }\textbf {\bibinfo {volume} {317}},\ \bibinfo {pages} {1911}
  (\bibinfo {year} {2007})}\BibitemShut {NoStop}%
\bibitem [{\citenamefont {Delaune}\ \emph {et~al.}(2012)\citenamefont
  {Delaune}, \citenamefont {Fran{\c{c}}ois}, \citenamefont {Shih},\ and\
  \citenamefont {Amacher}}]{delaune12}%
  \BibitemOpen
  \bibfield  {author} {\bibinfo {author} {\bibfnamefont {E.~A.}\ \bibnamefont
  {Delaune}}, \bibinfo {author} {\bibfnamefont {P.}~\bibnamefont
  {Fran{\c{c}}ois}}, \bibinfo {author} {\bibfnamefont {N.~P.}\ \bibnamefont
  {Shih}}, \ and\ \bibinfo {author} {\bibfnamefont {S.~L.}\ \bibnamefont
  {Amacher}},\ }\href@noop {} {\bibfield  {journal} {\bibinfo  {journal} {Dev.
  Cell}\ }\textbf {\bibinfo {volume} {23}},\ \bibinfo {pages} {995} (\bibinfo
  {year} {2012})}\BibitemShut {NoStop}%
\bibitem [{\citenamefont {Shih}\ \emph {et~al.}(2015)\citenamefont {Shih},
  \citenamefont {Fran{\c{c}}ois}, \citenamefont {Delaune},\ and\ \citenamefont
  {Amacher}}]{shih15}%
  \BibitemOpen
  \bibfield  {author} {\bibinfo {author} {\bibfnamefont {N.~P.}\ \bibnamefont
  {Shih}}, \bibinfo {author} {\bibfnamefont {P.}~\bibnamefont
  {Fran{\c{c}}ois}}, \bibinfo {author} {\bibfnamefont {E.~A.}\ \bibnamefont
  {Delaune}}, \ and\ \bibinfo {author} {\bibfnamefont {S.~L.}\ \bibnamefont
  {Amacher}},\ }\href@noop {} {\bibfield  {journal} {\bibinfo  {journal}
  {Development}\ }\textbf {\bibinfo {volume} {142}},\ \bibinfo {pages} {1785}
  (\bibinfo {year} {2015})}\BibitemShut {NoStop}%
\bibitem [{\citenamefont {Rohde}\ \emph {et~al.}(2024)\citenamefont {Rohde},
  \citenamefont {Bercowsky-Rama}, \citenamefont {Valentin}, \citenamefont
  {Naganathan}, \citenamefont {Desai}, \citenamefont {Strnad}, \citenamefont
  {Soroldoni},\ and\ \citenamefont {Oates}}]{rohde24}%
  \BibitemOpen
  \bibfield  {author} {\bibinfo {author} {\bibfnamefont {L.~A.}\ \bibnamefont
  {Rohde}}, \bibinfo {author} {\bibfnamefont {A.}~\bibnamefont
  {Bercowsky-Rama}}, \bibinfo {author} {\bibfnamefont {G.}~\bibnamefont
  {Valentin}}, \bibinfo {author} {\bibfnamefont {S.~R.}\ \bibnamefont
  {Naganathan}}, \bibinfo {author} {\bibfnamefont {R.~A.}\ \bibnamefont
  {Desai}}, \bibinfo {author} {\bibfnamefont {P.}~\bibnamefont {Strnad}},
  \bibinfo {author} {\bibfnamefont {D.}~\bibnamefont {Soroldoni}}, \ and\
  \bibinfo {author} {\bibfnamefont {A.~C.}\ \bibnamefont {Oates}},\ }\href@noop
  {} {\bibfield  {journal} {\bibinfo  {journal} {eLife}\ }\textbf {\bibinfo
  {volume} {13}},\ \bibinfo {pages} {RP93764} (\bibinfo {year}
  {2024})}\BibitemShut {NoStop}%
\bibitem [{\citenamefont {Soroldoni}\ \emph {et~al.}(2014)\citenamefont
  {Soroldoni}, \citenamefont {J{\"o}rg}, \citenamefont {Morelli}, \citenamefont
  {Richmond}, \citenamefont {Schindelin}, \citenamefont {J{\"u}licher},\ and\
  \citenamefont {Oates}}]{soroldoni14}%
  \BibitemOpen
  \bibfield  {author} {\bibinfo {author} {\bibfnamefont {D.}~\bibnamefont
  {Soroldoni}}, \bibinfo {author} {\bibfnamefont {D.~J.}\ \bibnamefont
  {J{\"o}rg}}, \bibinfo {author} {\bibfnamefont {L.~G.}\ \bibnamefont
  {Morelli}}, \bibinfo {author} {\bibfnamefont {D.~L.}\ \bibnamefont
  {Richmond}}, \bibinfo {author} {\bibfnamefont {J.}~\bibnamefont
  {Schindelin}}, \bibinfo {author} {\bibfnamefont {F.}~\bibnamefont
  {J{\"u}licher}}, \ and\ \bibinfo {author} {\bibfnamefont {A.~C.}\
  \bibnamefont {Oates}},\ }\href@noop {} {\bibfield  {journal} {\bibinfo
  {journal} {Science}\ }\textbf {\bibinfo {volume} {345}},\ \bibinfo {pages}
  {222} (\bibinfo {year} {2014})}\BibitemShut {NoStop}%
\bibitem [{\citenamefont {Eck}\ \emph {et~al.}(2024)\citenamefont {Eck},
  \citenamefont {Moretti}, \citenamefont {Schlomann}, \citenamefont
  {Bragantini}, \citenamefont {Lange}, \citenamefont {Zhao}, \citenamefont
  {VijayKumar}, \citenamefont {Valentin}, \citenamefont {Loureiro},
  \citenamefont {Soroldoni}, \citenamefont {Royer}, \citenamefont {Oates},\
  and\ \citenamefont {Garcia}}]{eck24}%
  \BibitemOpen
  \bibfield  {author} {\bibinfo {author} {\bibfnamefont {E.}~\bibnamefont
  {Eck}}, \bibinfo {author} {\bibfnamefont {B.}~\bibnamefont {Moretti}},
  \bibinfo {author} {\bibfnamefont {B.~H.}\ \bibnamefont {Schlomann}}, \bibinfo
  {author} {\bibfnamefont {J.}~\bibnamefont {Bragantini}}, \bibinfo {author}
  {\bibfnamefont {M.}~\bibnamefont {Lange}}, \bibinfo {author} {\bibfnamefont
  {X.}~\bibnamefont {Zhao}}, \bibinfo {author} {\bibfnamefont {S.}~\bibnamefont
  {VijayKumar}}, \bibinfo {author} {\bibfnamefont {G.}~\bibnamefont
  {Valentin}}, \bibinfo {author} {\bibfnamefont {C.}~\bibnamefont {Loureiro}},
  \bibinfo {author} {\bibfnamefont {D.}~\bibnamefont {Soroldoni}}, \bibinfo
  {author} {\bibfnamefont {L.~A.}\ \bibnamefont {Royer}}, \bibinfo {author}
  {\bibfnamefont {A.~C.}\ \bibnamefont {Oates}}, \ and\ \bibinfo {author}
  {\bibfnamefont {H.~G.}\ \bibnamefont {Garcia}},\ }\href {\doibase doi:
  10.1101/2024.01.03.574108.} {\bibfield  {journal} {\bibinfo  {journal}
  {bioRxiv}\ } (\bibinfo {year} {2024}),\ doi:
  10.1101/2024.01.03.574108.}\BibitemShut {Stop}%
\bibitem [{\citenamefont {Oginuma}\ \emph {et~al.}(2010)\citenamefont
  {Oginuma}, \citenamefont {Takahashi}, \citenamefont {Kitajima}, \citenamefont
  {Kiso}, \citenamefont {Kanno}, \citenamefont {Kimura},\ and\ \citenamefont
  {Saga}}]{oginuma10}%
  \BibitemOpen
  \bibfield  {author} {\bibinfo {author} {\bibfnamefont {M.}~\bibnamefont
  {Oginuma}}, \bibinfo {author} {\bibfnamefont {Y.}~\bibnamefont {Takahashi}},
  \bibinfo {author} {\bibfnamefont {S.}~\bibnamefont {Kitajima}}, \bibinfo
  {author} {\bibfnamefont {M.}~\bibnamefont {Kiso}}, \bibinfo {author}
  {\bibfnamefont {J.}~\bibnamefont {Kanno}}, \bibinfo {author} {\bibfnamefont
  {A.}~\bibnamefont {Kimura}}, \ and\ \bibinfo {author} {\bibfnamefont
  {Y.}~\bibnamefont {Saga}},\ }\href@noop {} {\bibfield  {journal} {\bibinfo
  {journal} {Development}\ }\textbf {\bibinfo {volume} {137}},\ \bibinfo
  {pages} {1515} (\bibinfo {year} {2010})}\BibitemShut {NoStop}%
\bibitem [{\citenamefont {Yabe}\ \emph {et~al.}(2023)\citenamefont {Yabe},
  \citenamefont {Uriu},\ and\ \citenamefont {Takada}}]{yabe23}%
  \BibitemOpen
  \bibfield  {author} {\bibinfo {author} {\bibfnamefont {T.}~\bibnamefont
  {Yabe}}, \bibinfo {author} {\bibfnamefont {K.}~\bibnamefont {Uriu}}, \ and\
  \bibinfo {author} {\bibfnamefont {S.}~\bibnamefont {Takada}},\ }\href@noop {}
  {\bibfield  {journal} {\bibinfo  {journal} {Nature Communications}\ }\textbf
  {\bibinfo {volume} {14}},\ \bibinfo {pages} {2115} (\bibinfo {year}
  {2023})}\BibitemShut {NoStop}%
\bibitem [{\citenamefont {Lawton}\ \emph {et~al.}(2013)\citenamefont {Lawton},
  \citenamefont {Nandi},\ and\ \citenamefont {Stulberg}}]{lawton13}%
  \BibitemOpen
  \bibfield  {author} {\bibinfo {author} {\bibfnamefont {A.}~\bibnamefont
  {Lawton}}, \bibinfo {author} {\bibfnamefont {A.}~\bibnamefont {Nandi}}, \
  and\ \bibinfo {author} {\bibfnamefont {M.}~\bibnamefont {Stulberg}},\
  }\href@noop {} {\bibfield  {journal} {\bibinfo  {journal} {Development}\
  }\textbf {\bibinfo {volume} {140}},\ \bibinfo {pages} {573} (\bibinfo {year}
  {2013})}\BibitemShut {NoStop}%
\bibitem [{\citenamefont {Mongera}\ \emph {et~al.}(2018)\citenamefont
  {Mongera}, \citenamefont {Rowghanian}, \citenamefont {Gustafson},
  \citenamefont {Shelton}, \citenamefont {Kealhofer}, \citenamefont {Carn},
  \citenamefont {Serwane}, \citenamefont {Lucio}, \citenamefont {Giammona},\
  and\ \citenamefont {Camp{\`a}s}}]{mongera18}%
  \BibitemOpen
  \bibfield  {author} {\bibinfo {author} {\bibfnamefont {A.}~\bibnamefont
  {Mongera}}, \bibinfo {author} {\bibfnamefont {P.}~\bibnamefont {Rowghanian}},
  \bibinfo {author} {\bibfnamefont {H.~J.}\ \bibnamefont {Gustafson}}, \bibinfo
  {author} {\bibfnamefont {E.}~\bibnamefont {Shelton}}, \bibinfo {author}
  {\bibfnamefont {D.~A.}\ \bibnamefont {Kealhofer}}, \bibinfo {author}
  {\bibfnamefont {E.~K.}\ \bibnamefont {Carn}}, \bibinfo {author}
  {\bibfnamefont {F.}~\bibnamefont {Serwane}}, \bibinfo {author} {\bibfnamefont
  {A.~A.}\ \bibnamefont {Lucio}}, \bibinfo {author} {\bibfnamefont
  {J.}~\bibnamefont {Giammona}}, \ and\ \bibinfo {author} {\bibfnamefont
  {O.}~\bibnamefont {Camp{\`a}s}},\ }\href@noop {} {\bibfield  {journal}
  {\bibinfo  {journal} {Nature}\ }\textbf {\bibinfo {volume} {561}},\ \bibinfo
  {pages} {401} (\bibinfo {year} {2018})}\BibitemShut {NoStop}%
\bibitem [{\citenamefont {Uriu}\ \emph {et~al.}(2021)\citenamefont {Uriu},
  \citenamefont {Liao}, \citenamefont {Oates},\ and\ \citenamefont
  {Morelli}}]{uriu21}%
  \BibitemOpen
  \bibfield  {author} {\bibinfo {author} {\bibfnamefont {K.}~\bibnamefont
  {Uriu}}, \bibinfo {author} {\bibfnamefont {B.-K.}\ \bibnamefont {Liao}},
  \bibinfo {author} {\bibfnamefont {A.~C.}\ \bibnamefont {Oates}}, \ and\
  \bibinfo {author} {\bibfnamefont {L.~G.}\ \bibnamefont {Morelli}},\
  }\href@noop {} {\bibfield  {journal} {\bibinfo  {journal} {Elife}\ }\textbf
  {\bibinfo {volume} {10}},\ \bibinfo {pages} {e61358} (\bibinfo {year}
  {2021})}\BibitemShut {NoStop}%
\bibitem [{\citenamefont {Pourqui{\'e}}(2011)}]{pourquie11}%
  \BibitemOpen
  \bibfield  {author} {\bibinfo {author} {\bibfnamefont {O.}~\bibnamefont
  {Pourqui{\'e}}},\ }\href@noop {} {\bibfield  {journal} {\bibinfo  {journal}
  {Cell}\ }\textbf {\bibinfo {volume} {145}},\ \bibinfo {pages} {650} (\bibinfo
  {year} {2011})}\BibitemShut {NoStop}%
\bibitem [{\citenamefont {Oates}\ \emph {et~al.}(2012)\citenamefont {Oates},
  \citenamefont {Morelli},\ and\ \citenamefont {Ares}}]{oates12}%
  \BibitemOpen
  \bibfield  {author} {\bibinfo {author} {\bibfnamefont {A.~C.}\ \bibnamefont
  {Oates}}, \bibinfo {author} {\bibfnamefont {L.~G.}\ \bibnamefont {Morelli}},
  \ and\ \bibinfo {author} {\bibfnamefont {S.}~\bibnamefont {Ares}},\
  }\href@noop {} {\bibfield  {journal} {\bibinfo  {journal} {Development}\
  }\textbf {\bibinfo {volume} {139}},\ \bibinfo {pages} {625} (\bibinfo {year}
  {2012})}\BibitemShut {NoStop}%
\bibitem [{\citenamefont {Hubaud}\ and\ \citenamefont
  {Pourqui{\'{e}}}(2014)}]{hubaud14}%
  \BibitemOpen
  \bibfield  {author} {\bibinfo {author} {\bibfnamefont {A.}~\bibnamefont
  {Hubaud}}\ and\ \bibinfo {author} {\bibfnamefont {O.}~\bibnamefont
  {Pourqui{\'{e}}}},\ }\href@noop {} {\bibfield  {journal} {\bibinfo  {journal}
  {Nat. Rev. Mol. Cell Biol.}\ }\textbf {\bibinfo {volume} {15}},\ \bibinfo
  {pages} {709} (\bibinfo {year} {2014})}\BibitemShut {NoStop}%
\bibitem [{\citenamefont {Ghosh}\ \emph {et~al.}(2022)\citenamefont {Ghosh},
  \citenamefont {Frasca}, \citenamefont {Rizzo}, \citenamefont {Majhi},
  \citenamefont {Rakshit}, \citenamefont {Alfaro-Bittner},\ and\ \citenamefont
  {Boccaletti}}]{ghosh2022}%
  \BibitemOpen
  \bibfield  {author} {\bibinfo {author} {\bibfnamefont {D.}~\bibnamefont
  {Ghosh}}, \bibinfo {author} {\bibfnamefont {M.}~\bibnamefont {Frasca}},
  \bibinfo {author} {\bibfnamefont {A.}~\bibnamefont {Rizzo}}, \bibinfo
  {author} {\bibfnamefont {S.}~\bibnamefont {Majhi}}, \bibinfo {author}
  {\bibfnamefont {S.}~\bibnamefont {Rakshit}}, \bibinfo {author} {\bibfnamefont
  {K.}~\bibnamefont {Alfaro-Bittner}}, \ and\ \bibinfo {author} {\bibfnamefont
  {S.}~\bibnamefont {Boccaletti}},\ }\href@noop {} {\bibfield  {journal}
  {\bibinfo  {journal} {Physics Reports}\ }\textbf {\bibinfo {volume} {949}},\
  \bibinfo {pages} {1} (\bibinfo {year} {2022})}\BibitemShut {NoStop}%
\bibitem [{\citenamefont {Frasca}\ \emph {et~al.}(2008)\citenamefont {Frasca},
  \citenamefont {Buscarino}, \citenamefont {Rizzo}, \citenamefont {Fortuna},\
  and\ \citenamefont {Boccaletti}}]{frasca08}%
  \BibitemOpen
  \bibfield  {author} {\bibinfo {author} {\bibfnamefont {M.}~\bibnamefont
  {Frasca}}, \bibinfo {author} {\bibfnamefont {A.}~\bibnamefont {Buscarino}},
  \bibinfo {author} {\bibfnamefont {A.}~\bibnamefont {Rizzo}}, \bibinfo
  {author} {\bibfnamefont {L.}~\bibnamefont {Fortuna}}, \ and\ \bibinfo
  {author} {\bibfnamefont {S.}~\bibnamefont {Boccaletti}},\ }\href@noop {}
  {\bibfield  {journal} {\bibinfo  {journal} {Physical Review Letters}\
  }\textbf {\bibinfo {volume} {100}},\ \bibinfo {pages} {044102} (\bibinfo
  {year} {2008})}\BibitemShut {NoStop}%
\bibitem [{\citenamefont {Uriu}\ \emph {et~al.}(2010)\citenamefont {Uriu},
  \citenamefont {Morishita},\ and\ \citenamefont {Iwasa}}]{uriu10a}%
  \BibitemOpen
  \bibfield  {author} {\bibinfo {author} {\bibfnamefont {K.}~\bibnamefont
  {Uriu}}, \bibinfo {author} {\bibfnamefont {Y.}~\bibnamefont {Morishita}}, \
  and\ \bibinfo {author} {\bibfnamefont {Y.}~\bibnamefont {Iwasa}},\
  }\href@noop {} {\bibfield  {journal} {\bibinfo  {journal} {Proc. Natl. Acad.
  Sci. USA}\ }\textbf {\bibinfo {volume} {107}},\ \bibinfo {pages} {4979}
  (\bibinfo {year} {2010})}\BibitemShut {NoStop}%
\bibitem [{\citenamefont {Fujiwara}\ \emph {et~al.}(2011)\citenamefont
  {Fujiwara}, \citenamefont {Kurths},\ and\ \citenamefont
  {D{\'{i}}az-Guilera}}]{fujiwara11}%
  \BibitemOpen
  \bibfield  {author} {\bibinfo {author} {\bibfnamefont {N.}~\bibnamefont
  {Fujiwara}}, \bibinfo {author} {\bibfnamefont {J.}~\bibnamefont {Kurths}}, \
  and\ \bibinfo {author} {\bibfnamefont {A.}~\bibnamefont
  {D{\'{i}}az-Guilera}},\ }\href@noop {} {\bibfield  {journal} {\bibinfo
  {journal} {Phys. Rev. E}\ }\textbf {\bibinfo {volume} {83}},\ \bibinfo
  {pages} {025101} (\bibinfo {year} {2011})}\BibitemShut {NoStop}%
\bibitem [{\citenamefont {Uriu}\ \emph {et~al.}(2013)\citenamefont {Uriu},
  \citenamefont {Ares}, \citenamefont {Oates},\ and\ \citenamefont
  {Morelli}}]{uriu13}%
  \BibitemOpen
  \bibfield  {author} {\bibinfo {author} {\bibfnamefont {K.}~\bibnamefont
  {Uriu}}, \bibinfo {author} {\bibfnamefont {S.}~\bibnamefont {Ares}}, \bibinfo
  {author} {\bibfnamefont {A.~C.}\ \bibnamefont {Oates}}, \ and\ \bibinfo
  {author} {\bibfnamefont {L.~G.}\ \bibnamefont {Morelli}},\ }\href@noop {}
  {\bibfield  {journal} {\bibinfo  {journal} {Phys. Rev. E}\ }\textbf {\bibinfo
  {volume} {87}},\ \bibinfo {pages} {032911} (\bibinfo {year}
  {2013})}\BibitemShut {NoStop}%
\bibitem [{\citenamefont {Majhi}\ and\ \citenamefont {Ghosh}(2017)}]{majhi17}%
  \BibitemOpen
  \bibfield  {author} {\bibinfo {author} {\bibfnamefont {S.}~\bibnamefont
  {Majhi}}\ and\ \bibinfo {author} {\bibfnamefont {D.}~\bibnamefont {Ghosh}},\
  }\href@noop {} {\bibfield  {journal} {\bibinfo  {journal} {Chaos: An
  Interdisciplinary Journal of Nonlinear Science}\ }\textbf {\bibinfo {volume}
  {27}},\ \bibinfo {pages} {053115} (\bibinfo {year} {2017})}\BibitemShut
  {NoStop}%
\bibitem [{\citenamefont {Levis}\ \emph {et~al.}(2017)\citenamefont {Levis},
  \citenamefont {Pagonabarraga},\ and\ \citenamefont
  {D{\'{i}}az-Guilera}}]{levis17}%
  \BibitemOpen
  \bibfield  {author} {\bibinfo {author} {\bibfnamefont {D.}~\bibnamefont
  {Levis}}, \bibinfo {author} {\bibfnamefont {I.}~\bibnamefont
  {Pagonabarraga}}, \ and\ \bibinfo {author} {\bibfnamefont {A.}~\bibnamefont
  {D{\'{i}}az-Guilera}},\ }\href@noop {} {\bibfield  {journal} {\bibinfo
  {journal} {Phys. Rev. X}\ }\textbf {\bibinfo {volume} {7}},\ \bibinfo {pages}
  {011028} (\bibinfo {year} {2017})}\BibitemShut {NoStop}%
\bibitem [{\citenamefont {Majhi}\ \emph {et~al.}(2019)\citenamefont {Majhi},
  \citenamefont {Ghosh},\ and\ \citenamefont {Kurths}}]{majhi19}%
  \BibitemOpen
  \bibfield  {author} {\bibinfo {author} {\bibfnamefont {S.}~\bibnamefont
  {Majhi}}, \bibinfo {author} {\bibfnamefont {D.}~\bibnamefont {Ghosh}}, \ and\
  \bibinfo {author} {\bibfnamefont {J.}~\bibnamefont {Kurths}},\ }\href
  {\doibase 10.1103/PhysRevE.99.012308} {\bibfield  {journal} {\bibinfo
  {journal} {Phys. Rev. E}\ }\textbf {\bibinfo {volume} {99}},\ \bibinfo
  {pages} {012308} (\bibinfo {year} {2019})}\BibitemShut {NoStop}%
\bibitem [{\citenamefont {Paulo}\ and\ \citenamefont
  {Tasinkevych}(2021)}]{paulo21}%
  \BibitemOpen
  \bibfield  {author} {\bibinfo {author} {\bibfnamefont {G.}~\bibnamefont
  {Paulo}}\ and\ \bibinfo {author} {\bibfnamefont {M.}~\bibnamefont
  {Tasinkevych}},\ }\href@noop {} {\bibfield  {journal} {\bibinfo  {journal}
  {Physical Review E}\ }\textbf {\bibinfo {volume} {104}},\ \bibinfo {pages}
  {014204} (\bibinfo {year} {2021})}\BibitemShut {NoStop}%
\bibitem [{\citenamefont {Li}\ and\ \citenamefont {Uchida}(2022)}]{li22}%
  \BibitemOpen
  \bibfield  {author} {\bibinfo {author} {\bibfnamefont {B.}~\bibnamefont
  {Li}}\ and\ \bibinfo {author} {\bibfnamefont {N.}~\bibnamefont {Uchida}},\
  }\href {\doibase 10.1103/PhysRevE.106.054210} {\bibfield  {journal} {\bibinfo
   {journal} {Phys. Rev. E}\ }\textbf {\bibinfo {volume} {106}},\ \bibinfo
  {pages} {054210} (\bibinfo {year} {2022})}\BibitemShut {NoStop}%
\bibitem [{\citenamefont {Gro{\ss}mann}\ \emph {et~al.}(2016)\citenamefont
  {Gro{\ss}mann}, \citenamefont {Peruani},\ and\ \citenamefont
  {B{\"a}r}}]{grossmann16}%
  \BibitemOpen
  \bibfield  {author} {\bibinfo {author} {\bibfnamefont {R.}~\bibnamefont
  {Gro{\ss}mann}}, \bibinfo {author} {\bibfnamefont {F.}~\bibnamefont
  {Peruani}}, \ and\ \bibinfo {author} {\bibfnamefont {M.}~\bibnamefont
  {B{\"a}r}},\ }\href@noop {} {\bibfield  {journal} {\bibinfo  {journal}
  {Physical Review E}\ }\textbf {\bibinfo {volume} {93}},\ \bibinfo {pages}
  {040102} (\bibinfo {year} {2016})}\BibitemShut {NoStop}%
\bibitem [{\citenamefont {Banerjee}\ and\ \citenamefont
  {Basu}(2017)}]{banerjee17}%
  \BibitemOpen
  \bibfield  {author} {\bibinfo {author} {\bibfnamefont {T.}~\bibnamefont
  {Banerjee}}\ and\ \bibinfo {author} {\bibfnamefont {A.}~\bibnamefont
  {Basu}},\ }\href@noop {} {\bibfield  {journal} {\bibinfo  {journal} {Physical
  Review E}\ }\textbf {\bibinfo {volume} {96}},\ \bibinfo {pages} {022201}
  (\bibinfo {year} {2017})}\BibitemShut {NoStop}%
\bibitem [{\citenamefont {Uriu}\ and\ \citenamefont {Morelli}(2014)}]{uriu14a}%
  \BibitemOpen
  \bibfield  {author} {\bibinfo {author} {\bibfnamefont {K.}~\bibnamefont
  {Uriu}}\ and\ \bibinfo {author} {\bibfnamefont {L.~G.}\ \bibnamefont
  {Morelli}},\ }\href@noop {} {\bibfield  {journal} {\bibinfo  {journal}
  {Biophys. J.}\ }\textbf {\bibinfo {volume} {107}},\ \bibinfo {pages} {514}
  (\bibinfo {year} {2014})}\BibitemShut {NoStop}%
\bibitem [{\citenamefont {Skufca}\ and\ \citenamefont
  {Bollt}(2004)}]{skufca04}%
  \BibitemOpen
  \bibfield  {author} {\bibinfo {author} {\bibfnamefont {J.~D.}\ \bibnamefont
  {Skufca}}\ and\ \bibinfo {author} {\bibfnamefont {E.~M.}\ \bibnamefont
  {Bollt}},\ }\href@noop {} {\bibfield  {journal} {\bibinfo  {journal}
  {Mathematical Biosciences and Engineering}\ }\textbf {\bibinfo {volume}
  {1}},\ \bibinfo {pages} {347} (\bibinfo {year} {2004})}\BibitemShut {NoStop}%
\bibitem [{\citenamefont {Zhou}\ \emph {et~al.}(2016)\citenamefont {Zhou},
  \citenamefont {Zou}, \citenamefont {Guan}, \citenamefont {Liu},\ and\
  \citenamefont {Boccaletti}}]{zhou16}%
  \BibitemOpen
  \bibfield  {author} {\bibinfo {author} {\bibfnamefont {J.}~\bibnamefont
  {Zhou}}, \bibinfo {author} {\bibfnamefont {Y.}~\bibnamefont {Zou}}, \bibinfo
  {author} {\bibfnamefont {S.}~\bibnamefont {Guan}}, \bibinfo {author}
  {\bibfnamefont {Z.}~\bibnamefont {Liu}}, \ and\ \bibinfo {author}
  {\bibfnamefont {S.}~\bibnamefont {Boccaletti}},\ }\href@noop {} {\bibfield
  {journal} {\bibinfo  {journal} {Scientific Reports}\ }\textbf {\bibinfo
  {volume} {6}},\ \bibinfo {pages} {35979} (\bibinfo {year}
  {2016})}\BibitemShut {NoStop}%
\bibitem [{\citenamefont {Anwar}\ \emph {et~al.}(2022)\citenamefont {Anwar},
  \citenamefont {Rakshit}, \citenamefont {Ghosh},\ and\ \citenamefont
  {Bollt}}]{anwar22}%
  \BibitemOpen
  \bibfield  {author} {\bibinfo {author} {\bibfnamefont {M.~S.}\ \bibnamefont
  {Anwar}}, \bibinfo {author} {\bibfnamefont {S.}~\bibnamefont {Rakshit}},
  \bibinfo {author} {\bibfnamefont {D.}~\bibnamefont {Ghosh}}, \ and\ \bibinfo
  {author} {\bibfnamefont {E.~M.}\ \bibnamefont {Bollt}},\ }\href@noop {}
  {\bibfield  {journal} {\bibinfo  {journal} {Physical Review E}\ }\textbf
  {\bibinfo {volume} {105}},\ \bibinfo {pages} {024303} (\bibinfo {year}
  {2022})}\BibitemShut {NoStop}%
\bibitem [{\citenamefont {Majhi}\ \emph {et~al.}(2022)\citenamefont {Majhi},
  \citenamefont {Rakshit},\ and\ \citenamefont {Ghosh}}]{majhi22}%
  \BibitemOpen
  \bibfield  {author} {\bibinfo {author} {\bibfnamefont {S.}~\bibnamefont
  {Majhi}}, \bibinfo {author} {\bibfnamefont {S.}~\bibnamefont {Rakshit}}, \
  and\ \bibinfo {author} {\bibfnamefont {D.}~\bibnamefont {Ghosh}},\
  }\href@noop {} {\bibfield  {journal} {\bibinfo  {journal} {Chaos: An
  Interdisciplinary Journal of Nonlinear Science}\ }\textbf {\bibinfo {volume}
  {32}},\ \bibinfo {pages} {042101} (\bibinfo {year} {2022})}\BibitemShut
  {NoStop}%
\bibitem [{\citenamefont {Tanaka}(2007)}]{tanaka07}%
  \BibitemOpen
  \bibfield  {author} {\bibinfo {author} {\bibfnamefont {D.}~\bibnamefont
  {Tanaka}},\ }\href@noop {} {\bibfield  {journal} {\bibinfo  {journal} {Phys.
  Rev. Lett.}\ }\textbf {\bibinfo {volume} {99}},\ \bibinfo {pages} {134103}
  (\bibinfo {year} {2007})}\BibitemShut {NoStop}%
\bibitem [{\citenamefont {O'Keeffe}\ \emph {et~al.}(2017)\citenamefont
  {O'Keeffe}, \citenamefont {Hong},\ and\ \citenamefont
  {Strogatz}}]{okeeffe17}%
  \BibitemOpen
  \bibfield  {author} {\bibinfo {author} {\bibfnamefont {K.~P.}\ \bibnamefont
  {O'Keeffe}}, \bibinfo {author} {\bibfnamefont {H.}~\bibnamefont {Hong}}, \
  and\ \bibinfo {author} {\bibfnamefont {S.~H.}\ \bibnamefont {Strogatz}},\
  }\href@noop {} {\bibfield  {journal} {\bibinfo  {journal} {Nat. Commun.}\
  }\textbf {\bibinfo {volume} {8}},\ \bibinfo {pages} {1504} (\bibinfo {year}
  {2017})}\BibitemShut {NoStop}%
\bibitem [{\citenamefont {Levis}\ \emph {et~al.}(2019)\citenamefont {Levis},
  \citenamefont {Pagonabarraga},\ and\ \citenamefont {Liebchen}}]{levis19}%
  \BibitemOpen
  \bibfield  {author} {\bibinfo {author} {\bibfnamefont {D.}~\bibnamefont
  {Levis}}, \bibinfo {author} {\bibfnamefont {I.}~\bibnamefont
  {Pagonabarraga}}, \ and\ \bibinfo {author} {\bibfnamefont {B.}~\bibnamefont
  {Liebchen}},\ }\href@noop {} {\bibfield  {journal} {\bibinfo  {journal}
  {Physical Review Research}\ }\textbf {\bibinfo {volume} {1}},\ \bibinfo
  {pages} {023026} (\bibinfo {year} {2019})}\BibitemShut {NoStop}%
\bibitem [{\citenamefont {Sar}\ \emph {et~al.}(2022)\citenamefont {Sar},
  \citenamefont {Chowdhury}, \citenamefont {Perc},\ and\ \citenamefont
  {Ghosh}}]{sar22}%
  \BibitemOpen
  \bibfield  {author} {\bibinfo {author} {\bibfnamefont {G.~K.}\ \bibnamefont
  {Sar}}, \bibinfo {author} {\bibfnamefont {S.~N.}\ \bibnamefont {Chowdhury}},
  \bibinfo {author} {\bibfnamefont {M.}~\bibnamefont {Perc}}, \ and\ \bibinfo
  {author} {\bibfnamefont {D.}~\bibnamefont {Ghosh}},\ }\href@noop {}
  {\bibfield  {journal} {\bibinfo  {journal} {New Journal of Physics}\ }\textbf
  {\bibinfo {volume} {24}},\ \bibinfo {pages} {043004} (\bibinfo {year}
  {2022})}\BibitemShut {NoStop}%
\bibitem [{\citenamefont {O'Keeffe}\ \emph {et~al.}(2022)\citenamefont
  {O'Keeffe}, \citenamefont {Ceron},\ and\ \citenamefont
  {Petersen}}]{okeeffe22}%
  \BibitemOpen
  \bibfield  {author} {\bibinfo {author} {\bibfnamefont {K.}~\bibnamefont
  {O'Keeffe}}, \bibinfo {author} {\bibfnamefont {S.}~\bibnamefont {Ceron}}, \
  and\ \bibinfo {author} {\bibfnamefont {K.}~\bibnamefont {Petersen}},\
  }\href@noop {} {\bibfield  {journal} {\bibinfo  {journal} {Physical Review
  E}\ }\textbf {\bibinfo {volume} {105}},\ \bibinfo {pages} {014211} (\bibinfo
  {year} {2022})}\BibitemShut {NoStop}%
\bibitem [{\citenamefont {Ceron}\ \emph {et~al.}(2023)\citenamefont {Ceron},
  \citenamefont {O’Keeffe},\ and\ \citenamefont {Petersen}}]{ceron23}%
  \BibitemOpen
  \bibfield  {author} {\bibinfo {author} {\bibfnamefont {S.}~\bibnamefont
  {Ceron}}, \bibinfo {author} {\bibfnamefont {K.}~\bibnamefont {O’Keeffe}}, \
  and\ \bibinfo {author} {\bibfnamefont {K.}~\bibnamefont {Petersen}},\
  }\href@noop {} {\bibfield  {journal} {\bibinfo  {journal} {Nature
  Communications}\ }\textbf {\bibinfo {volume} {14}},\ \bibinfo {pages} {940}
  (\bibinfo {year} {2023})}\BibitemShut {NoStop}%
\bibitem [{\citenamefont {Fulton}\ \emph {et~al.}(2022)\citenamefont {Fulton},
  \citenamefont {Verd},\ and\ \citenamefont {Steventon}}]{fulton22}%
  \BibitemOpen
  \bibfield  {author} {\bibinfo {author} {\bibfnamefont {T.}~\bibnamefont
  {Fulton}}, \bibinfo {author} {\bibfnamefont {B.}~\bibnamefont {Verd}}, \ and\
  \bibinfo {author} {\bibfnamefont {B.}~\bibnamefont {Steventon}},\ }\href@noop
  {} {\bibfield  {journal} {\bibinfo  {journal} {Royal Society Open Science}\
  }\textbf {\bibinfo {volume} {9}},\ \bibinfo {pages} {211293} (\bibinfo {year}
  {2022})}\BibitemShut {NoStop}%
\bibitem [{\citenamefont {Uriu}\ and\ \citenamefont {Morelli}(2017)}]{uriu17a}%
  \BibitemOpen
  \bibfield  {author} {\bibinfo {author} {\bibfnamefont {K.}~\bibnamefont
  {Uriu}}\ and\ \bibinfo {author} {\bibfnamefont {L.~G.}\ \bibnamefont
  {Morelli}},\ }\href@noop {} {\bibfield  {journal} {\bibinfo  {journal}
  {Development, growth \& differentiation}\ } (\bibinfo {year}
  {2017})}\BibitemShut {NoStop}%
\bibitem [{\citenamefont {Petrungaro}\ \emph {et~al.}(2017)\citenamefont
  {Petrungaro}, \citenamefont {Uriu},\ and\ \citenamefont
  {Morelli}}]{petrungaro17}%
  \BibitemOpen
  \bibfield  {author} {\bibinfo {author} {\bibfnamefont {G.}~\bibnamefont
  {Petrungaro}}, \bibinfo {author} {\bibfnamefont {K.}~\bibnamefont {Uriu}}, \
  and\ \bibinfo {author} {\bibfnamefont {L.~G.}\ \bibnamefont {Morelli}},\
  }\href@noop {} {\bibfield  {journal} {\bibinfo  {journal} {Phys. Rev. E}\
  }\textbf {\bibinfo {volume} {96}},\ \bibinfo {pages} {062210} (\bibinfo
  {year} {2017})}\BibitemShut {NoStop}%
\bibitem [{\citenamefont {Petrungaro}\ \emph {et~al.}(2019)\citenamefont
  {Petrungaro}, \citenamefont {Uriu},\ and\ \citenamefont
  {Morelli}}]{petrungaro19}%
  \BibitemOpen
  \bibfield  {author} {\bibinfo {author} {\bibfnamefont {G.}~\bibnamefont
  {Petrungaro}}, \bibinfo {author} {\bibfnamefont {K.}~\bibnamefont {Uriu}}, \
  and\ \bibinfo {author} {\bibfnamefont {L.~G.}\ \bibnamefont {Morelli}},\
  }\href@noop {} {\bibfield  {journal} {\bibinfo  {journal} {Physical Review
  E}\ }\textbf {\bibinfo {volume} {99}},\ \bibinfo {pages} {062207} (\bibinfo
  {year} {2019})}\BibitemShut {NoStop}%
\bibitem [{\citenamefont {Kuramoto}(1984)}]{kuramoto}%
  \BibitemOpen
  \bibfield  {author} {\bibinfo {author} {\bibfnamefont {Y.}~\bibnamefont
  {Kuramoto}},\ }\href@noop {} {\emph {\bibinfo {title} {Chemical Oscillations,
  Waves, and Turbulence.}}}\ (\bibinfo  {publisher} {Springer-Verlag},\
  \bibinfo {address} {Berlin},\ \bibinfo {year} {1984})\BibitemShut {NoStop}%
\bibitem [{\citenamefont {Wiley}\ \emph {et~al.}(2006)\citenamefont {Wiley},
  \citenamefont {Strogatz},\ and\ \citenamefont {Girvan}}]{wiley06}%
  \BibitemOpen
  \bibfield  {author} {\bibinfo {author} {\bibfnamefont {D.~A.}\ \bibnamefont
  {Wiley}}, \bibinfo {author} {\bibfnamefont {S.~H.}\ \bibnamefont {Strogatz}},
  \ and\ \bibinfo {author} {\bibfnamefont {M.}~\bibnamefont {Girvan}},\
  }\href@noop {} {\bibfield  {journal} {\bibinfo  {journal} {Chaos}\ }\textbf
  {\bibinfo {volume} {16}},\ \bibinfo {pages} {015103} (\bibinfo {year}
  {2006})}\BibitemShut {NoStop}%
\bibitem [{\citenamefont {Peruani}\ \emph {et~al.}(2010)\citenamefont
  {Peruani}, \citenamefont {Nicola},\ and\ \citenamefont
  {Morelli}}]{peruani10}%
  \BibitemOpen
  \bibfield  {author} {\bibinfo {author} {\bibfnamefont {F.}~\bibnamefont
  {Peruani}}, \bibinfo {author} {\bibfnamefont {E.~M.}\ \bibnamefont {Nicola}},
  \ and\ \bibinfo {author} {\bibfnamefont {L.~G.}\ \bibnamefont {Morelli}},\
  }\href@noop {} {\bibfield  {journal} {\bibinfo  {journal} {New J. Phys.}\
  }\textbf {\bibinfo {volume} {12}},\ \bibinfo {pages} {093029} (\bibinfo
  {year} {2010})}\BibitemShut {NoStop}%
\bibitem [{\citenamefont {Gardiner}(2009)}]{gardiner09}%
  \BibitemOpen
  \bibfield  {author} {\bibinfo {author} {\bibfnamefont {C.}~\bibnamefont
  {Gardiner}},\ }\href@noop {} {\emph {\bibinfo {title} {Stochastic
  methods}}},\ Vol.~\bibinfo {volume} {4}\ (\bibinfo  {publisher} {Springer
  Berlin},\ \bibinfo {year} {2009})\BibitemShut {NoStop}%
\bibitem [{\citenamefont {Strogatz}\ and\ \citenamefont
  {Mirollo}(1991)}]{strogatz91}%
  \BibitemOpen
  \bibfield  {author} {\bibinfo {author} {\bibfnamefont {S.~H.}\ \bibnamefont
  {Strogatz}}\ and\ \bibinfo {author} {\bibfnamefont {R.~E.}\ \bibnamefont
  {Mirollo}},\ }\href@noop {} {\bibfield  {journal} {\bibinfo  {journal}
  {Journal of Statistical Physics}\ }\textbf {\bibinfo {volume} {63}},\
  \bibinfo {pages} {613} (\bibinfo {year} {1991})}\BibitemShut {NoStop}%
\bibitem [{\citenamefont {J{\"{o}}rg}\ \emph {et~al.}(2015)\citenamefont
  {J{\"{o}}rg}, \citenamefont {Morelli}, \citenamefont {Soroldoni},
  \citenamefont {Oates},\ and\ \citenamefont {J{\"{u}}licher}}]{jorg15}%
  \BibitemOpen
  \bibfield  {author} {\bibinfo {author} {\bibfnamefont {D.~J.}\ \bibnamefont
  {J{\"{o}}rg}}, \bibinfo {author} {\bibfnamefont {L.~G.}\ \bibnamefont
  {Morelli}}, \bibinfo {author} {\bibfnamefont {D.}~\bibnamefont {Soroldoni}},
  \bibinfo {author} {\bibfnamefont {A.~C.}\ \bibnamefont {Oates}}, \ and\
  \bibinfo {author} {\bibfnamefont {F.}~\bibnamefont {J{\"{u}}licher}},\
  }\href@noop {} {\bibfield  {journal} {\bibinfo  {journal} {New J. Phys.}\
  }\textbf {\bibinfo {volume} {17}},\ \bibinfo {pages} {093042} (\bibinfo
  {year} {2015})}\BibitemShut {NoStop}%
\bibitem [{\citenamefont {Morelli}\ \emph {et~al.}(2009)\citenamefont
  {Morelli}, \citenamefont {Ares}, \citenamefont {Herrgen}, \citenamefont
  {Schr{\"{o}}ter}, \citenamefont {J{\"{u}}licher},\ and\ \citenamefont
  {Oates}}]{morelli09}%
  \BibitemOpen
  \bibfield  {author} {\bibinfo {author} {\bibfnamefont {L.~G.}\ \bibnamefont
  {Morelli}}, \bibinfo {author} {\bibfnamefont {S.}~\bibnamefont {Ares}},
  \bibinfo {author} {\bibfnamefont {L.}~\bibnamefont {Herrgen}}, \bibinfo
  {author} {\bibfnamefont {C.}~\bibnamefont {Schr{\"{o}}ter}}, \bibinfo
  {author} {\bibfnamefont {F.}~\bibnamefont {J{\"{u}}licher}}, \ and\ \bibinfo
  {author} {\bibfnamefont {A.~C.}\ \bibnamefont {Oates}},\ }\href@noop {}
  {\bibfield  {journal} {\bibinfo  {journal} {HFSP}\ }\textbf {\bibinfo
  {volume} {3}},\ \bibinfo {pages} {55} (\bibinfo {year} {2009})}\BibitemShut
  {NoStop}%
\bibitem [{\citenamefont {Herrgen}\ \emph {et~al.}(2010)\citenamefont
  {Herrgen}, \citenamefont {Ares}, \citenamefont {Morelli}, \citenamefont
  {Schr{\"{o}}ter}, \citenamefont {J{\"{u}}licher},\ and\ \citenamefont
  {Oates}}]{herrgen10}%
  \BibitemOpen
  \bibfield  {author} {\bibinfo {author} {\bibfnamefont {L.}~\bibnamefont
  {Herrgen}}, \bibinfo {author} {\bibfnamefont {S.}~\bibnamefont {Ares}},
  \bibinfo {author} {\bibfnamefont {L.~G.}\ \bibnamefont {Morelli}}, \bibinfo
  {author} {\bibfnamefont {C.}~\bibnamefont {Schr{\"{o}}ter}}, \bibinfo
  {author} {\bibfnamefont {F.}~\bibnamefont {J{\"{u}}licher}}, \ and\ \bibinfo
  {author} {\bibfnamefont {A.~C.}\ \bibnamefont {Oates}},\ }\href@noop {}
  {\bibfield  {journal} {\bibinfo  {journal} {Curr. Biol.}\ }\textbf {\bibinfo
  {volume} {20}},\ \bibinfo {pages} {1244} (\bibinfo {year}
  {2010})}\BibitemShut {NoStop}%
\bibitem [{\citenamefont {Giudicelli}\ \emph {et~al.}(2007)\citenamefont
  {Giudicelli}, \citenamefont {Ozbudak}, \citenamefont {Wright},\ and\
  \citenamefont {Lewis}}]{giudicelli07}%
  \BibitemOpen
  \bibfield  {author} {\bibinfo {author} {\bibfnamefont {F.}~\bibnamefont
  {Giudicelli}}, \bibinfo {author} {\bibfnamefont {E.~M.}\ \bibnamefont
  {Ozbudak}}, \bibinfo {author} {\bibfnamefont {G.~J.}\ \bibnamefont {Wright}},
  \ and\ \bibinfo {author} {\bibfnamefont {J.}~\bibnamefont {Lewis}},\
  }\href@noop {} {\bibfield  {journal} {\bibinfo  {journal} {PLoS Biol}\
  }\textbf {\bibinfo {volume} {5}},\ \bibinfo {pages} {e150} (\bibinfo {year}
  {2007})}\BibitemShut {NoStop}%
\bibitem [{\citenamefont {Nakamasu}\ \emph {et~al.}(2009)\citenamefont
  {Nakamasu}, \citenamefont {Takahashi}, \citenamefont {Kanbe},\ and\
  \citenamefont {Kondo}}]{nakamasu09}%
  \BibitemOpen
  \bibfield  {author} {\bibinfo {author} {\bibfnamefont {A.}~\bibnamefont
  {Nakamasu}}, \bibinfo {author} {\bibfnamefont {G.}~\bibnamefont {Takahashi}},
  \bibinfo {author} {\bibfnamefont {A.}~\bibnamefont {Kanbe}}, \ and\ \bibinfo
  {author} {\bibfnamefont {S.}~\bibnamefont {Kondo}},\ }\href@noop {}
  {\bibfield  {journal} {\bibinfo  {journal} {Proceedings of the National
  Academy of Sciences}\ }\textbf {\bibinfo {volume} {106}},\ \bibinfo {pages}
  {8429} (\bibinfo {year} {2009})}\BibitemShut {NoStop}%
\bibitem [{\citenamefont {B{\'e}naz{\'e}raf}\ \emph {et~al.}(2010)\citenamefont
  {B{\'e}naz{\'e}raf}, \citenamefont {Francois}, \citenamefont {Baker},
  \citenamefont {Denans}, \citenamefont {Little},\ and\ \citenamefont
  {Pourqui{\'e}}}]{benazeraf10}%
  \BibitemOpen
  \bibfield  {author} {\bibinfo {author} {\bibfnamefont {B.}~\bibnamefont
  {B{\'e}naz{\'e}raf}}, \bibinfo {author} {\bibfnamefont {P.}~\bibnamefont
  {Francois}}, \bibinfo {author} {\bibfnamefont {R.~E.}\ \bibnamefont {Baker}},
  \bibinfo {author} {\bibfnamefont {N.}~\bibnamefont {Denans}}, \bibinfo
  {author} {\bibfnamefont {C.~D.}\ \bibnamefont {Little}}, \ and\ \bibinfo
  {author} {\bibfnamefont {O.}~\bibnamefont {Pourqui{\'e}}},\ }\href@noop {}
  {\bibfield  {journal} {\bibinfo  {journal} {Nature}\ }\textbf {\bibinfo
  {volume} {466}},\ \bibinfo {pages} {248} (\bibinfo {year}
  {2010})}\BibitemShut {NoStop}%
\bibitem [{\citenamefont {Delfini}\ \emph {et~al.}(2005)\citenamefont
  {Delfini}, \citenamefont {Dubrulle}, \citenamefont {Malapert}, \citenamefont
  {Chal},\ and\ \citenamefont {Pourqui{\'{e}}}}]{delfini05}%
  \BibitemOpen
  \bibfield  {author} {\bibinfo {author} {\bibfnamefont {M.~C.}\ \bibnamefont
  {Delfini}}, \bibinfo {author} {\bibfnamefont {J.}~\bibnamefont {Dubrulle}},
  \bibinfo {author} {\bibfnamefont {P.}~\bibnamefont {Malapert}}, \bibinfo
  {author} {\bibfnamefont {J.}~\bibnamefont {Chal}}, \ and\ \bibinfo {author}
  {\bibfnamefont {O.}~\bibnamefont {Pourqui{\'{e}}}},\ }\href@noop {}
  {\bibfield  {journal} {\bibinfo  {journal} {Proc. Natl. Acad. Sci. USA}\
  }\textbf {\bibinfo {volume} {102}},\ \bibinfo {pages} {11343} (\bibinfo
  {year} {2005})}\BibitemShut {NoStop}%
\bibitem [{\citenamefont {Banavar}\ \emph {et~al.}(2021)\citenamefont
  {Banavar}, \citenamefont {Carn}, \citenamefont {Rowghanian}, \citenamefont
  {Stooke-Vaughan}, \citenamefont {Kim},\ and\ \citenamefont
  {Camp{\`a}s}}]{banavar21}%
  \BibitemOpen
  \bibfield  {author} {\bibinfo {author} {\bibfnamefont {S.~P.}\ \bibnamefont
  {Banavar}}, \bibinfo {author} {\bibfnamefont {E.~K.}\ \bibnamefont {Carn}},
  \bibinfo {author} {\bibfnamefont {P.}~\bibnamefont {Rowghanian}}, \bibinfo
  {author} {\bibfnamefont {G.}~\bibnamefont {Stooke-Vaughan}}, \bibinfo
  {author} {\bibfnamefont {S.}~\bibnamefont {Kim}}, \ and\ \bibinfo {author}
  {\bibfnamefont {O.}~\bibnamefont {Camp{\`a}s}},\ }\href@noop {} {\bibfield
  {journal} {\bibinfo  {journal} {Scientific Reports}\ }\textbf {\bibinfo
  {volume} {11}},\ \bibinfo {pages} {1} (\bibinfo {year} {2021})}\BibitemShut
  {NoStop}%
\end{thebibliography}%

\end{document}